\documentclass[usenatbib]{mnras}
\usepackage[utf8]{inputenc}
\usepackage{amsmath}
\usepackage{graphicx}
\usepackage[T1]{fontenc}
\usepackage{ae,aecompl}
\usepackage{subfig}

\newcommand{\cm}{{\mathrm{cm}}}

\newcommand{\s}{{\mathrm{s}}}
\newcommand{\cmsqps}{\cm^2\s^{-1}}
\newcommand{\pc}{{\mathrm{pc}}}

\newcommand{\Msun}{\mathrm{M}_{\odot}}

\usepackage[inline,ignoremode]{trackchanges}

\title[The phase structure of cosmic ray driven outflows in stream fed disc galaxies]{The phase structure of cosmic ray driven outflows in stream fed disc galaxies}

\author[N. Peschken, M. Hanasz, T. Naab, D. W\'{o}lta\'{n}ski, A. Gawryszczak]{N. Peschken$^{1}$\thanks{Contact e-mail:
    \href{mailto:npeschken@umk.pl}{npeschken@umk.pl}}, M. Hanasz$^{1}$, T. Naab$^{2}$, D. W\'{o}lta\'{n}ski$^{1}$, A. Gawryszczak$^{3}$\\
$^{1}$ Institute of Astronomy, Faculty of Physics, Astronomy and Informatics, Nicolaus Copernicus University,\\
ul. Grudziadzka 5/7, PL-87-100 Toruń, Poland\\
$^{2}$ Max Planck Institute for Astrophysics, Karl-Schwarzschild-Str. 1, 85748 Garching, Germany\\
$^{3}$Nicolaus Copernicus Astronomical Center, Bartycka 18, 00-716 Warsaw, Poland
}

\pubyear{2022}

\begin{document}
\label{firstpage}
\pagerange{\pageref{firstpage}--\pageref{lastpage}}
\maketitle

\begin{abstract}
Feeding with gas in streams is predicted to be an important galaxy growth mechanism. Using an idealised setup, we study the impact of stream feeding (with 10$^7$~M$_{\odot}$~Myr$^{-1}$ rate) on the star formation and outflows of disc galaxies with $\sim$10$^{11}$~M$_{\odot}$ baryonic mass. The magneto-hydrodynamical simulations are carried out with the \textsc{piernik} code and include star formation, feedback from supernova, and cosmic ray advection and diffusion. We find that stream accretion enhances galactic star formation. Lower angular momentum streams result in more compact discs, higher star formation rates and stronger outflows. In agreement with previous studies, models including cosmic rays launch stronger outflows travelling much further into the galactic halo. Cosmic ray supported outflows are also cooler than supernova only driven outflows. With cosmic rays, the star formation is suppressed and the thermal pressure is reduced. We find evidence for two distinct outflow phases. The warm outflows have high angular momentum and stay close to the galactic disc, while the hot outflow phase has low angular momentum and escapes from the centre deep into the halo. Cosmic rays can therefore have a strong impact on galaxy evolution by removing low angular momentum, possibly metal enriched gas from the disc and injecting it into the circumgalactic medium.
\end{abstract}
 
\begin{keywords}
galaxies: evolution -- galaxies: kinematics and dynamics -- galaxies: spiral -- galaxies: structure
\end{keywords}

\section{Introduction}

Gas outflows play an important role in galaxy evolution, being able to launch large scale galactic winds, trigger galactic fountains, remove gas from star forming regions and bring gas from the disc into the circumgalactic medium. Observations have shown them to be ubiquitous in galaxies \citep{1990ApJS...74..833H,1999ApJ...513..156M,2010ApJ...717..289S,2014ApJ...794..156R}. Understanding their origin and the mechanisms driving them is therefore a key question for galactic dynamics, and this topic is still actively investigated in galaxy models today. Feedback processes seem promising candidates to explain their formation, by injecting energy into the gas in AGN or star forming regions. In particular, thermal supernova feedback has shown capable of driving powerful outflows \citep{2006ApJ...653.1266J,2014A&A...570A..81H,2016MNRAS.456.3432G}, and some thermal driven winds have indeed been observed in galaxies \citep{1985Natur.317...44C}. However, in many cases, thermal feedback alone fails at reproducing a number of observables such as cooler outflowing gas \citep{2014ApJ...792....8W} or lower star formation \citep{2018MNRAS.475.3511G,2019MNRAS.485.3317S}, and often struggles to produce sufficiently strong outflows. The need for an additional non-thermal feedback mechanism has therefore arisen, and this has led to the increasing implementation of cosmic ray feedback in recent years (see \citealt{2017ARA&A..55...59N} and \citealt {2021LRCA....7....2H} for reviews). \\ 
Cosmic Rays are charged particles emitted in supernova remnants that are accelerated by diffusive supernova shocks to relativistic energies \citep{1977DoSSR.234.1306K,1978ApJ...221L..29B,2004MNRAS.353..550B}, and travel large distances via diffusion and advection \citep{2007ARNPS..57..285S,2013PhPl...20e5501Z,2015ARA&A..53..199G}. They can reach high energy densities in star forming regions, leading to equipartition with thermal gas \citep{1990ApJ...365..544B}, which, coupled with their ability to transfer energy and momentum to the gas, makes them an important component to be included in the ISM. They have indeed been shown to accelerate gas out of galactic discs, launching large scale outflows by creating a gradient of pressure pointing towards the disc \citep{1991A&A...245...79B, 2012MNRAS.423.2374U, 2013ApJ...777L..38H, 2013ApJ...777L..16B, 2014ApJ...797L..18S, 2016ApJ...824L..30P,2019A&A...630A.107D,2022arXiv220312029T,2022MNRAS.510.3917G}. The implementation of comic rays in supernova feedback has led to interesting results for the outflows, as compared to purely thermal feedback. The cosmic ray pressure indeed seems to push colder gas into the outflows \citep{2013ApJ...777L..16B,2014MNRAS.437.3312S,2018MNRAS.479.3042G}, to launch winds to higher altitudes \citep{2018ApJ...868..108B,2021MNRAS.501.3640H}, to suppress star formation \citep{2008A&A...481...33J,2017MNRAS.465.4500P,2021ApJ...910..126S}, to thicken the disc \citep{2014MNRAS.437.3312S, 2016ApJ...816L..19G}, and to trigger fountain mechanisms \citep{2018ApJ...856..112F}. However the importance of cosmic rays relative to thermal feedback is still debated, some studies finding that cosmic rays enhance outflows \citep{2014MNRAS.437.3312S,2016ApJ...816L..19G,2020A&A...638A.123D,2022MNRAS.513.5000F}, while some do not (\citealt{2013ApJ...777L..16B} for their MW-size model, \citealt{2020MNRAS.497.2623J}, \citealt{2021MNRAS.501.3640H} in the disc vicinity). The results also seem to highly depend on the galaxies mass and diffusion coefficients \citep{2014MNRAS.437.3312S,2019MNRAS.488.3716C,2020MNRAS.492.3465H}, and overall the feedback scheme used. Further research is therefore needed to fully understand cosmic ray driven outflows, and their importance and effects as compared to thermal driven ones. \\
In this context, we propose to investigate in this paper outflows generated in the presence of cosmic rays and thermal feedback, using MHD simulations of disc galaxies run with the \textsc{piernik} code. After showing for the first time in \cite{2021MNRAS.508.4269P} the link between angular momentum and cosmic ray driven outflows, we extend our study here in the presence of thermal feedback, and examine the composition of the outflows in terms of both temperature and angular momentum. Following the prediction from cosmological simulations that stream feeding is an important factor in galaxy growth \citep{2009Natur.457..451D}, we model the accretion by the disc of an external inflow of gas from the intergalactic medium with varying angular momentum, and observe the resulting generated outflows. Stream accretion has indeed been linked to star forming rings in high redshift galaxies \citep{2008ApJ...687...59G,2012ApJ...745...11G,2015MNRAS.449.2087D,2020MNRAS.496.5372D}, and can extend star formation activity by continuously feeding cold gas to the disc. \\
We organise the paper as follows. After first introducing and describing our simulations in section \ref{presentation}, we present our results in section \ref{results}, with the investigation of outflow production and evolution in the presence and absence of cosmic rays, the effect of varying the angular momentum of the stream, and the temperature and angular momentum composition of the outflows. We then discuss the results in section \ref{discussion}, and conclude in section \ref{ccl}. \\

     
\begin{figure*}
  \includegraphics[scale=0.50, trim=70 40 0 65, clip]{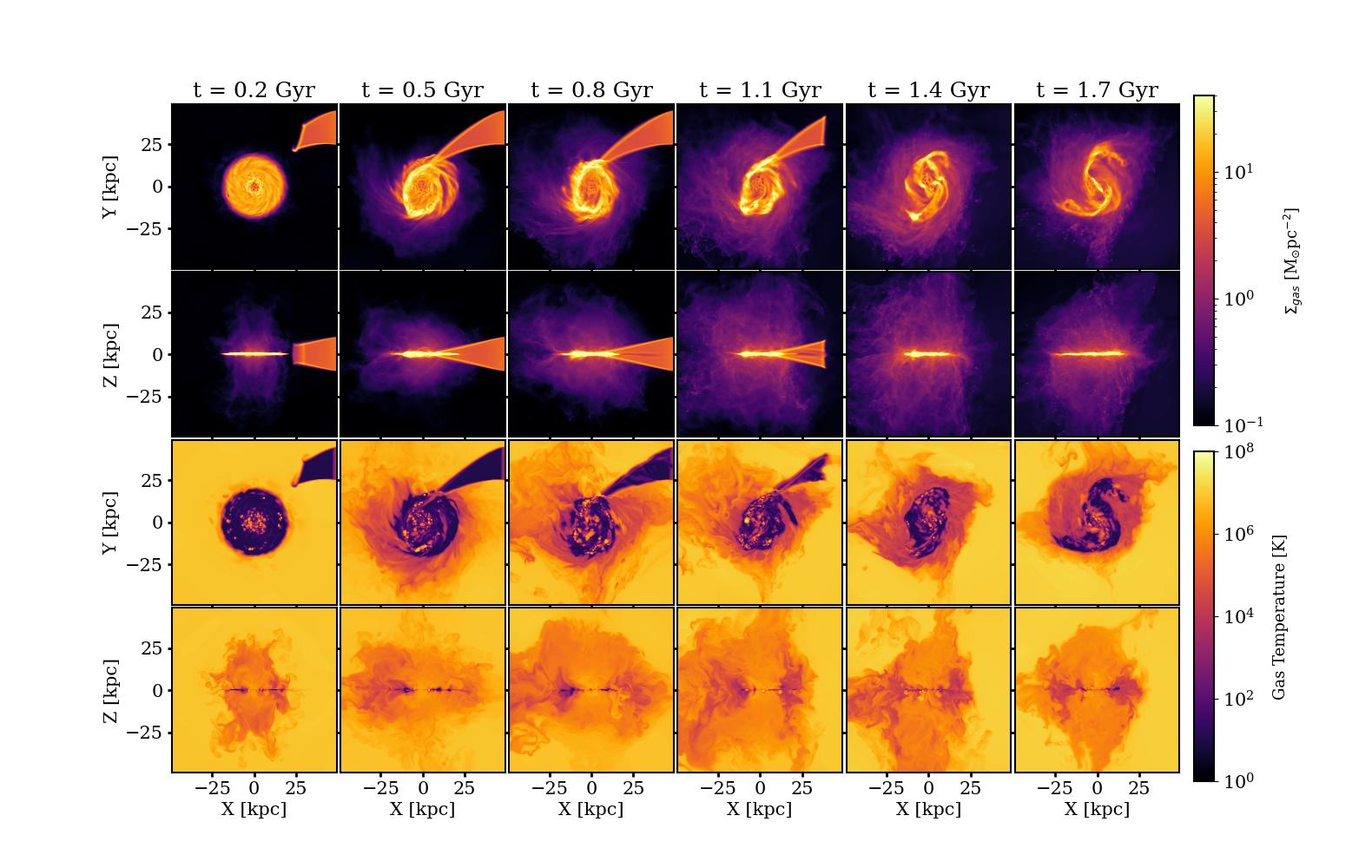}
\caption{Evolution of the cr+t3 simulation (CR + thermal feedback) in gas density and gas temperature, shown at different times (evolving from left to right). From top to bottom: face-one and edge-on views of the column density, and face-on and edge-on views of the temperature taken in a slice.}
  \label{dens_temp_evol}
\end{figure*}

\section{Presentation of the simulations}
\label{presentation}

   \subsection{The PIERNIK code}
   \label{piernik}

   We run our simulations with the grid-MHD \textsc{piernik} code \citep{2010EAS....42..275H, 2010EAS....42..281H, 2012EAS....56..363H, 2012EAS....56..367H}.
Multiple fluids such as thermal gas, dust and cosmic rays (CRs) can be simulated in \textsc{piernik}, as separate fluids interacting within the fluid approximation \citep{2003A&A...404..389H}. We give below a brief description of the code, and refer the reader to \cite{2013ApJ...777L..38H}, as well as the series of papers mentioned above, for more details. \\
The \textsc{piernik} code has been recently equipped with the HLLD Riemann solver by \cite{2005JCoPh.208..315M} combined with the \cite{2002JCoPh.175..645D} conservative formulation of MHD equations, which uses hyperbolic divergence cleaning. Time integration is performed with second order Runge-Kutta schemes.
We have tested the MHD algorithm performing various standard MHD tests. Results of two example tests are shown in Appendix \ref{sect:appendix}.
The Riemann solver, serving as an alternative to the previously used simple and robust but more diffusive Relaxing Total Variation Diminishing (RTVD) scheme \citep{jin-xin-95,2003PASP..115..303T,2003ApJS..149..447P}, has been used in simulations conducted for this paper.
\textsc{piernik} is available from the public git repository linked to the code webpage: \href{http://piernik.umk.pl}{piernik.umk.pl}. \\
The Poisson equation is solved by an iterative multigrid solver \citep{doi:10.1137/S1064827598346235}, with the potential at the domain boudaries handled by a multipole solver based on \citet{1977JCoPh..25...71J}.
CRs propagate with the diffusion-advection transport equation:

\begin{equation}
\frac{\partial e_{cr}}{\partial t} + \pmb{\nabla}(e_{cr}\pmb{v}) = \pmb{\nabla}(K\pmb{\nabla}e_{cr}) - p_{cr}(\pmb{\nabla}\cdot\pmb{v}) + Q
\end{equation}

\noindent with $e_{cr}$ being the cosmic ray energy density, $K$ the diffusivity coefficient, $p_{cr}=(\gamma_{cr}-1)e_{cr}$ the cosmic ray pressure (we use $\gamma_{cr}=4/3$), $\pmb{v}$ the gas velocity, and $Q$ the cosmic ray source term from supernova remnants.
We take a uniform diffusion coefficient of $K=9 \times 10^{28}\cmsqps$, in agreement with recent work on the impact of different coefficients on galaxy evolution and outflows (\citealt{2016ApJ...816L..19G, 2016MNRAS.456..582S, 2016ApJ...827L..29S, 2018MNRAS.479.3042G,2019MNRAS.488.3716C, 2019A&A...632A..12S, 2020MNRAS.492.3465H, 2022MNRAS.510.1184Q}) finding the most realistic values to be of the order of $10^{29} \cmsqps$, and also generating galactic scale outflows. \\
In contrast to \cite{2021MNRAS.508.4269P}, we set up isotropic CR diffusion in this paper to mitigate an excess of CR energy density with respect to magnetic and kinetic energy densities observed in our simulations. This effect, resulting supposedly from insufficient spatial resolution (cell sizes comparable to ${(200\pc)^3}$), may cause an excessive magnetic diffusivity and numerical reconnection smoothing out small-scale magnetic field structures. At higher resolution the small-scale turbulent fields would permit diffusion of CRs across the direction of dominant magnetic field component. By assuming isotropic diffusivity of CRs we compensate for the lack of small-scale magnetic field fluctuations that otherwise would lead to an easier escape of cosmic rays in the  direction perpendicular to the dominating horizontal configuration of large-scale magnetic field in the disc. \\
We recently added Nbody particles to \textsc{piernik} in order to model stars and dark matter, which we described first in \cite{2021MNRAS.508.4269P}. The corresponding particle mesh module includes a leapfrog scheme, and particle density is projected on the grid using a Triangular Shaped Cloud (TSC) scheme. The TSC algorithm is then used again to derive the particles accelerations after computation of the gravitational potential with the Poisson solver. The particle solver proved successful as we find a realistic evolution of the stellar disc staying thin over time and developping spirals, both in the simulations presented in \cite{2021MNRAS.508.4269P}, and in this paper (see Fig.~\ref{evolpart}). For this paper we implemented a new addition to the particle module allowing the formation of new stellar particles at each timestep, on top of the pre-existing stellar disc. We detail this process in section \ref{thermal}, describing our star formation recipe.\\

\begin{figure*}
  \includegraphics[scale=0.6, trim=50 40 0 70, clip]{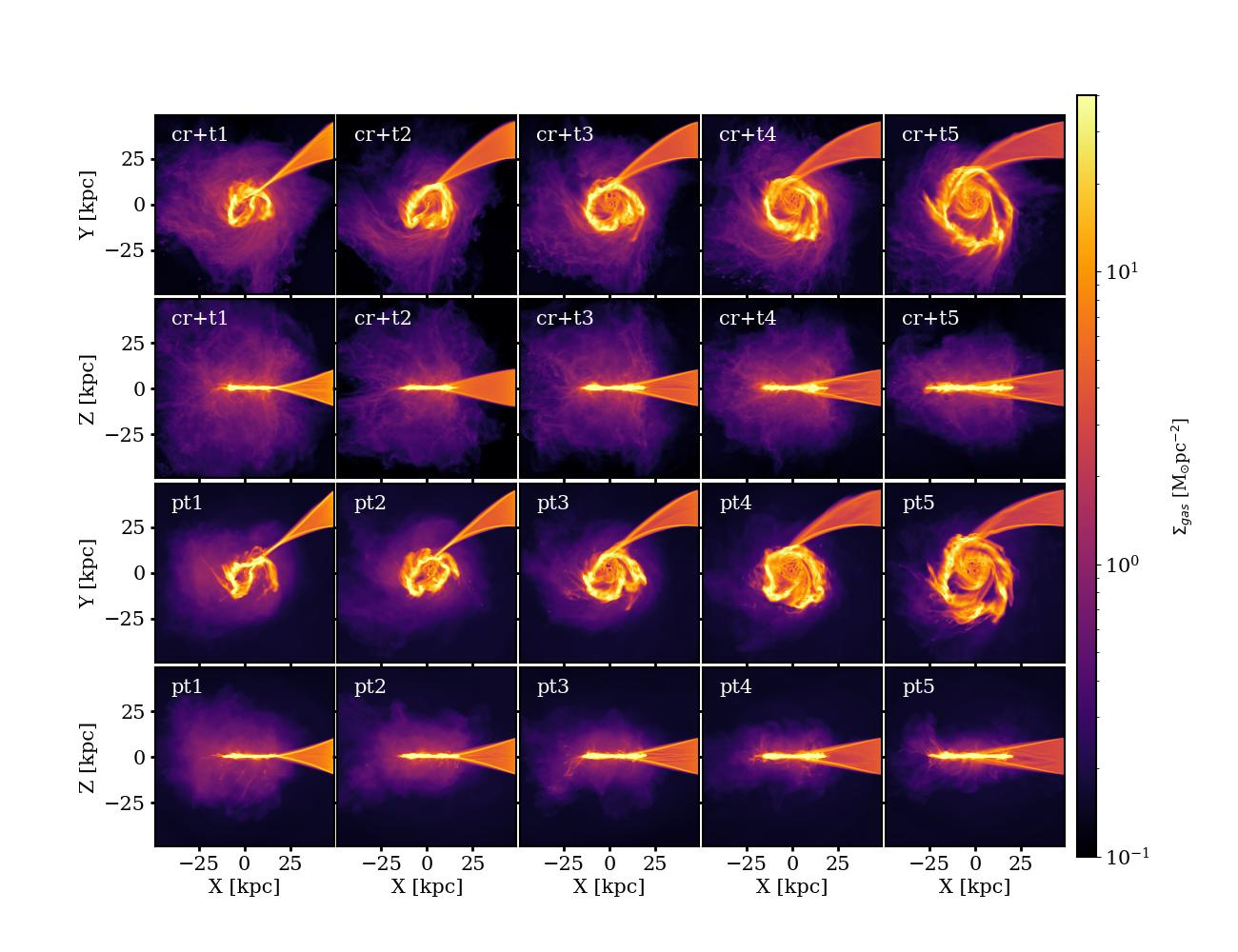}
\caption{Top two rows: face-on and edge-on display of the gas column density for the five simulations with both cosmic rays and thermal feedback (cr+t), at t = 1~Gyr. Bottom two rows: same for the cases without cosmic ray feedback, i.e. the pure thermal simulations (pt). From left to right the stream inflow feeds the disc with more angular momentum, resulting in different morphologies and outflow productions.}
\label{setup}
\end{figure*}

   \subsection{Galaxy initial conditions}
   \label{disc}
       
   In order to explore the impact of stream acretion on the outflows in disc galaxies, we use a setup similar to \cite{2021MNRAS.508.4269P}, which we summarize below. An exponential stellar disc (with a hyperbolic secant to model its vertical distribution) is built together with a dark matter halo from Nbody particles, using the BUILDGAL software \citep{1993ApJS...86..389H}. The disc has a mass of 8.56 $\times$ 10$^{10}$~M$_{\odot}$, a 1.6~kpc scale-height, a 5.3~kpc scale-length with a 79.5~kpc cut-off radius, a Toomre parameter of 1.5, and is composed of half a million particles. Dark matter is distributed in an isothermal halo of also half a million particles, of 4.28 $\times$ 10$^{11}$~M$_{\odot}$ mass and with a core of 5.3~kpc radius, as well as a 265~kpc cut-off radius. The gravitational potential corresponding to these two Nbody components is then derived in \textsc{piernik}, which will be used to build a gaseous disc in hydrostatic equilibrium with the stellar and dark matter distributions. The magnetic field is introduced in pressure equipartition with the gas, with an initial toroidal strength of 9.6~$\mu$G. We keep the same resolution as in \cite{2021MNRAS.508.4269P}, i.e. 512$^3$ cells in a (100~kpc)$^3$ volume with the disc placed in the center, and run our simulations for 1.7 Gyr. Cosmic rays are not present initially as they will be introduced via star formation (see section  \ref{thermal} below). \\
   One significant new addition to this setup is the introduction of cooling and heating of the thermal gas, in contrast to \cite{2021MNRAS.508.4269P}. The cooling of the gas is introduced using the Exact Integration Scheme \citep{2009ApJS..181..391T}, which provides an exact analytical solution of the gas temperature for a given cooling function. We take the cooling table from \cite{2020MNRAS.497.4857P}, which we fit using a piece-wise power law. The resulting piece-wise power law function can then be integrated in the Exact Integration Scheme of our new cooling module. At each time-step and in every grid cell independently, the gas temperature is first updated from the local internal energy; then the cooling time is derived from the cooling power-law relevant for this gas cell, and finally the new temperature is derived from the exact solution provided by the Exact Integration Scheme applied to this power-law. Heating of the gas, on the other hand, is mostly created by thermal energy injection (as we describe in the section below), and by shockwaves. A minimum temperature of 10~K is set, and a maximum of 10$^8$~K. The disc is initially set at a temperature of 10$^4$~K, while the hot halo is initialized with 8.3 $\times$ 10$^6$~K.\\

   \subsection{Stellar formation and feedback scheme}
   \label{thermal}

   We changed the feedback implementation in the \textsc{piernik} code. While in previous work modelizing galaxies with \textsc{piernik} \citep{2013ApJ...777L..38H,2021MNRAS.508.4269P}, the stellar feedback was purely in the form of cosmic rays injected in star forming areas, here we implemented a new scheme including both CR and thermal feedbacks, alongside stellar particles creation. We detail this scheme below. \\
   We select areas of star formation based on two criteria. At each timestep, a gas cell will form stars only if: \\
   (1) its density is greater than a given threshold density $\rho_{\mathrm{thr}}$ = 0.035~$\Msun$~$\pc^{-3}$ (which represents 1 Hydrogen atom~cm$^{-3}$ + 1 Helium atom every 10 H atoms). \\
   (2) its temperature is lower than $10^4$~K.\\
   If these two criteria are fulfilled, a fraction of this cell's gas content will be considered as star forming. The corresponding star formation rate (SFR) is derived with:
\begin{equation}
\rho_{SFR} = \epsilon \sqrt{\frac{G\rho^3}{32\pi}}
\end{equation}
   
\begin{figure*}
  \includegraphics[scale=0.59, trim=50 0 0 20, clip]{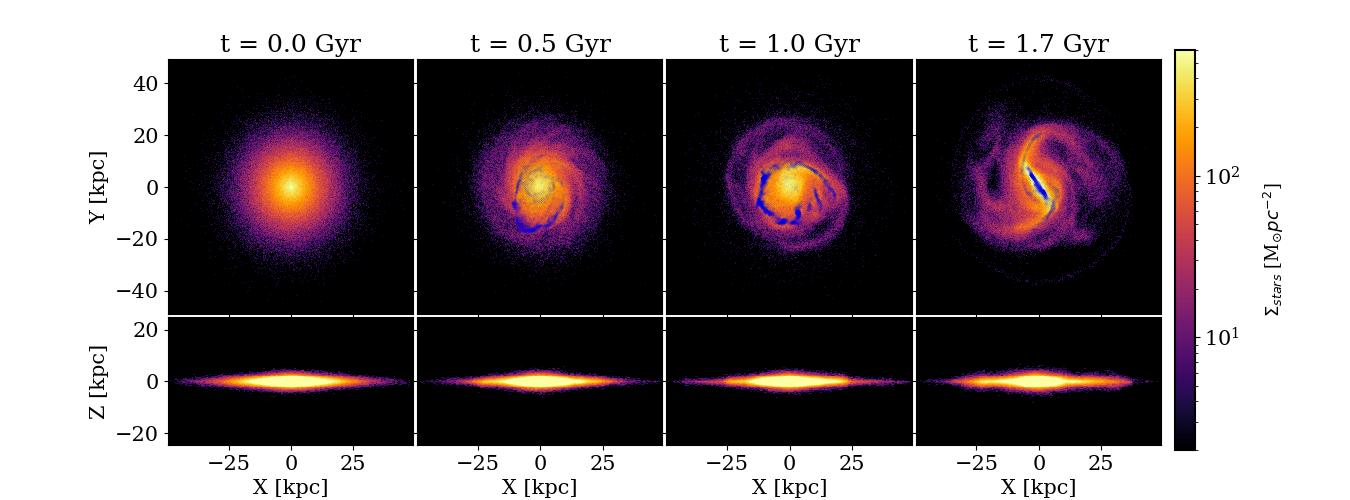}
\caption{Stellar projected density for the cr+t3 (CR + thermal) simulation seen face-on (top row) and edge-on (bottom row) at t = 0, 0.5 Gyr, 1 Gyr and 1.7~Gyr. New stellar particles (born in the last 100 Myr) are highlighted in blue.}
\label{evolpart}
\end{figure*}
 \noindent  $\epsilon$ being the SFR efficiency, set here to 10\%.
   This allows to derive the star forming mass in a given cell during one timestep, which will be converted into stars. This is done by removing the corresponding mass from the gas cell, and adding it to a stellar particle present in this cell. Stellar particles can therefore have a mass increasing over time, until they reach a maximum of $10^5$~$M_{\odot}$, at which point their mass is locked and cannot increase anymore, to avoid having too massive particles. If a gas cell is forming stars and does not contain a stellar particle, or only contains stellar particles with locked masses, a new star particle is created at the centre of this cell with the star forming mass. We note that the particles of the initial stellar disc are already locked with 1.72 $\times$ 10$^5$~$M_{\odot}$. \\
   Once a new stellar particle reaches its maximum mass of $10^5$~$M_{\odot}$, the gas cell in which it currently resides will release a stellar feedback energy corresponding to a thousand supernova explosions. This makes approximately for one supernova explosion (SNe) every 100~$M_{\odot}$. The reason we wait to reach the mass corresponding to 1000 SNe to initiate the stellar feedback is to avoid overcooling of the SNe which happens in dense environments \citep{1992ApJ...391..502K}. To further prevent too rapid cooling, we also implement a momentum injection in the gas cells surrounding the star forming cell prior to the supernova energy injection, representing dispersion of the gas in giant molecular cloud before the SNe by radiation pressure \citep{2005ApJ...618..569M} and stellar winds \citep{2002ApJ...566..302M}. We model this momentum feedback using the recipe of \cite{2013ApJ...770...25A} for momentum kicks, i.e. deposing at each time-step the momentum $p_{kick}$ isotropically in the 26 neighbouring cells for a duration of $t_k$ = 6.5~Myr, with $p_{kick}$ following Equation (4) of \cite{2013ApJ...770...25A}: \\
   \begin{equation}
p_{kick} = M_* p_1 \left(\frac{Z_*}{p_2}\right)^{p_3} \frac{dt}{t_k} \mathrm{\ \ g\ cm\ s}^{-1}
   \end{equation}
  \noindent with $M_*$ the initial stellar particle mass, $Z_*$ the stellar metallicity in units of solar metallicity and set to 1, $p_1$ = 1.8 $\times$ 10$^{40}$ g cm s$^{-1}$ M$_{\odot}^{-1}$, $p_2$ = 0.50, $p_3$ = 0.38 and $dt$ the time-step. \\
   After this initial 6.5~Myr of momentum feedback, the energy feedback corresponding to these 1000 SNe is injected in the star forming gas cell, over one timestep, in the form of CRs and/or thermal energy. We assume the energy of one SNe to be of $10^{51}$~erg, so that the energy injected for one stellar particle is of $10^{54}$~erg (1000 SNe). For the comparison of outflows between cosmic ray and thermal feedback, we run 2 types of simulations: simulations with pure thermal feedback (no CR energy injection) called $pt$, and simulations including both cosmic ray and thermal feedback, called $cr$+$t$. In the former, all the energy is thermal, while in the latter 90\% of this energy is injected in the form of thermal energy, and 10\% as CR energy.


   \subsection{Gas stream accretion with varying angular momentum}
   \label{stream}

   To simulate the stream feeding of the galaxy, and study the impact of inflows with different angular momentum, we proceed similarly to \cite{2021MNRAS.508.4269P} by introducing an inflow of gas providing angular momentum to the disc. We summarize this setup briefly here and refer the reader to \cite{2021MNRAS.508.4269P} for more details. The stream arrives from the YZ edge of our domain (Z being the direction perpendicular to the disc) corresponding to X = 50~kpc, and at Z = 0~kpc and Y = 35~kpc. This stream is attributed a constant negative velocity in the X direction (towards the inside of the domain), which we vary from one simulation to another to change the amount of angular momentum added to the disc. We however keep a constant gas flux in each case of 10$^7$~M$_{\odot}$~Myr$^{-1}$, by changing the density of the stream accordingly. The stream is maintained for a period of 1 Gyr from the start of the simulation (for a total of 10$^{10}$~M$_{\odot}$ mass input), and is fully accreted by the disc by t$\sim$1.2~Gyr. This leaves about 500~Myr for the disc to settle after being fed by the stream. The gas in the stream is cold (starting with 10$^4$~K, but cooling down before reaching the disc), in agreement with stream feeding predictions \citep{2009Natur.457..451D}, however star formation is disabled within the stream. Fig. \ref{dens_temp_evol} displays the evolution of the gas in the galaxy, and provides an illustration of this stream setup. Note that in contrast to \cite{2021MNRAS.508.4269P}, we here simulate a wider stream of (20~kpc)$^2$ cross-section, in order to have a more dilute and less violent inflow entry into the disc. We also change and extend the range of values taken for the X-velocities (and therefore for the added angular momentum): we now take inflow velocities ranging from 15 km/s to 150 km/s, instead of the previous 75 km/s maximum. We simulate here five different values of the inflow velocity: -15 km/s, -45 km/s, -75 km/s, -110 km/s, and -150 km/s, labelled accordingly 1,2,3,4 and 5. This leads to 10 simulations in total: pt1-5 (pure thermal feedbacks), and cr+t1-5 (CR + thermal feedback). As will be shown in this paper, changing the inflow velocity affects the galaxy evolution dramatically; we display in Fig. \ref{setup} the gas column density of the five cr+t simulations and the five pt simulations at t = 1~Gyr, where differences are already very visible. \\
   Videos displaying the gas density and temperature evolutions of all the simulations are available on the \textsc{piernik} webpage: \href{http://piernik.umk.pl/results/2022a/}{piernik.umk.pl/results/2022a}.

\begin{figure}
  \includegraphics[scale=0.5, trim=20 15 0 10, clip]{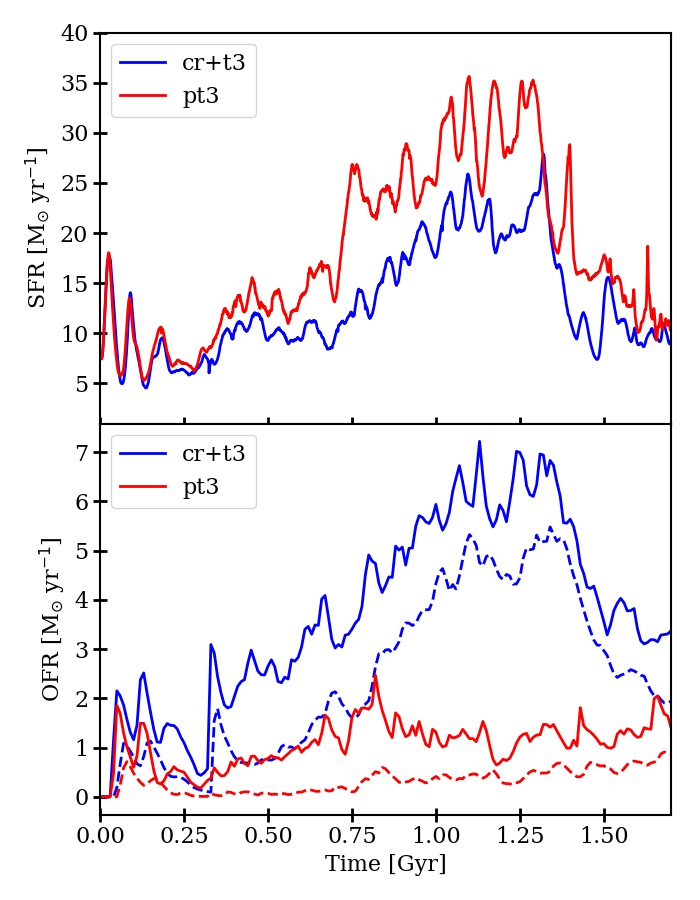}
  \caption{Comparison of the star formation rate (SFR, upper panel) and of the outflow rate (OFR, lower panel) for the simulations cr+t3 (CR+thermal) and pt3 (pure thermal). The outflow rate at 10 kpc is plotted with a solid line, and the one at 20 kpc with a dotted line.}
  \label{outf_sfr_tcr}
\end{figure}


   \begin{figure}
  \includegraphics[scale=0.53, trim=4 20 0 60, clip]{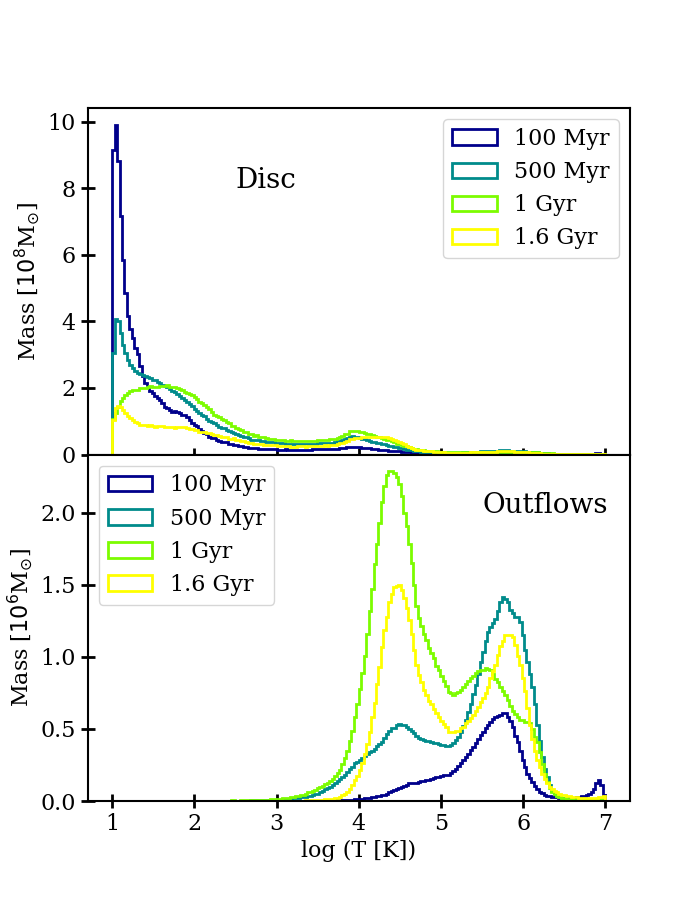}
\caption{Temperature distribution evolution in the disc (top panel), and in the outflows taken at 5~kpc altitude (lower panel), for the cr+t3 simulation (CR + thermal). For each time the gas temperatures are averaged over 200 Myr, and the outflows are taken within 30~kpc radius (corresponding to the disc). Bin size: 0.03 temperature logarithmic units. Outflows feed predominantly from the warm and hot gas of the disc.}
  \label{T_hist_evol}
   \end{figure}

\section{Results}
\label{results}

\begin{figure*}
  \includegraphics[scale=0.50, trim=70 40 0 65, clip]{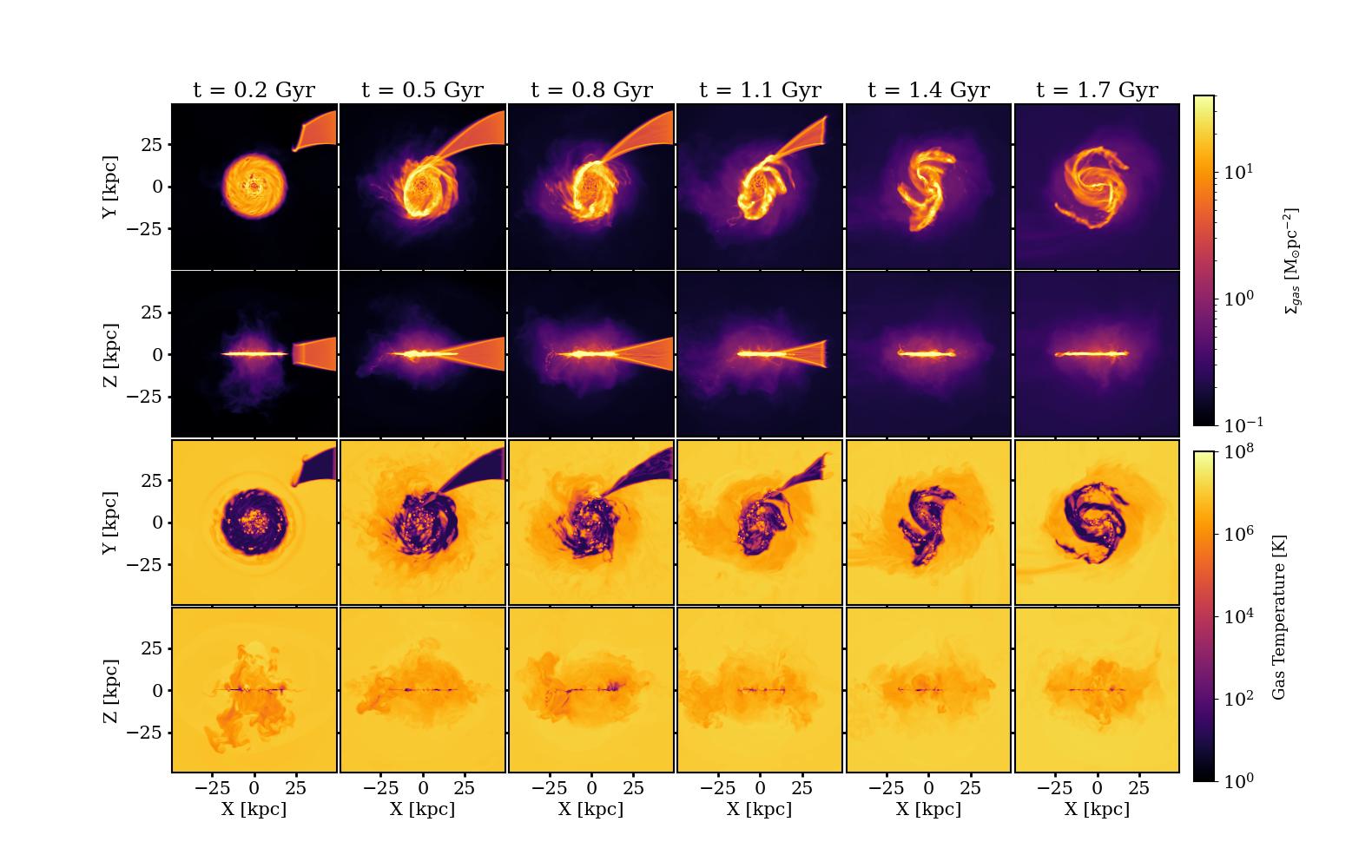}
\caption{Evolution of the gas column density and temperature, similarly to Fig. \ref{dens_temp_evol}, but displayed here for the pt3 simulation (pure thermal feedback).}
  \label{dens_temp_evol_pt3}
\end{figure*}

   \subsection{Galaxy evolution and outflows creation}
   \label{cr+t3}

We start by analysing a fiducial simulation, which includes both cosmic ray and thermal stellar feedback, and with a stream inflowing at -75~km/s X-velocity, i.e. the cr+t3 simulation. This simulation is  interesting because it has an intermediate value of angular momentum stream input in our set of simulations, but it is important to note that all the qualitative results presented in this section are also true for the other four cr+t simulations. \\
Looking at the galaxy stellar evolution in Fig. \ref{evolpart}, we see that from a smooth disc distribution, the galaxy quickly develops a spiral structure, and a strong bar towards the end of the simulation. Most of the new born stars are found in the spirals and the bar, which appear to be strong star formation areas. \\
We then turn to the gas column density evolution in Fig. \ref{dens_temp_evol} (top two rows). The stream is accreted smoothly by the galaxy, forming a ring (in agreement with previous studies such as \citealt{2020MNRAS.496.5372D}) of about 15~kpc radius, and eventually leading to the formation of spiral arms and even a bar. In the edge-on view, we can clearly see gas leaving the disc for the halo over time, showing the presence of significant outflows up to high altitudes above and below the disc. \\
To characterize the gas outflows observed in the edge-on view, we display in Fig. \ref{outf_sfr_tcr} (in blue) the outflow rate (OFR) at 10~kpc altitude as a function of time (computed as the gas flux crossing the 10~kpc vertical planes both above and below the disc), as well as the star formation rate (SFR). We already showed in \cite{2021MNRAS.508.4269P} the tight link between SFR and outflow production in the presence of CRs: as the SFR increases, more CRs are produced as a result, which quickly leave the disc and create a stronger vertical gradient of pressure (also shown in \citealt{2022ApJ...929..170A}) accelerating gas out of the disc into powerful outflows. We will investigate in details the role of thermal feedback in the outflow production in next section (section \ref{CR_therm}), but it is already interesting to note that despite the use of a new star formation recipe including the introduction of thermal energy injection in SNe, we still find that the OFR for cr+t3 is following the SFR closely over time. In the first $\sim$200 Myr, both show rapid variations, corresponding to initial star formation bursts. From around 300~Myr however, as the gas stream reaches the disc, both the SFR and the OFR start showing a more steady increase. This is a consequence of the stream feeding the gaseous disc: the gas density is locally increased, allowing for more star formation. It is also very visible that once the stream stops feeding the disc, around t = 1.3~Gyr, both the SFR and OFR drop by a factor $\sim$2. The OFR reaches values of $\sim$7 M$_{\odot}$ yr$^{-1}$, for a mass loading of about 0.3. This value fits within the range of mass loadings found by previous works on cosmic ray driven outflows (e.g. \citealt{2017ApJ...834..208R, 2018MNRAS.475..570J}). \\
Since one of the new additions to our simulations is the presence of gas cooling and heating, let us look at the gas temperature distribution in Fig. \ref{dens_temp_evol} (bottom two rows). In the disc region the high gas density leads to rapid cooling and very low temperatures, but with the notable presence of supernova explosions heating it locally. Once the spirals and the bar develop, the gas constituting them stays very cold, while the inter-arm region and the periphery of the disc shows warmer temperatures. Looking at the histogram of temperatures in the disc for different times (Fig. \ref{T_hist_evol}, top panel), we see that indeed initially the gas temperatures are mostly distributed close to the minimum allowed (10~K), but over time a second warmer phase develops around 2 $\times$ 10$^4$~K. This is most likely due to the heating by the thermal energy injected in supernovae explosions. The very cold gas fraction tends to decrease over time, as gas is heated from the SNe. Note that a very small fraction of the gas even reaches hotter temperatures close to 10$^6$~K. \\
   The gas present in the outflows in the edge-on view of Fig. \ref{dens_temp_evol} (bottom row) shows intermediate temperatures, being cooler than the halo but hotter than the disc. We display the evolution of the temperatures distribution in the outflows in Fig. \ref{T_hist_evol} (lower panel) for gas outflowing at 5~kpc altitude, so right above (and below) the disc. This evolution is interesting; the outflows start mostly as a hot phase at almost 10$^6$~K, but then develop a cooler phase around 10$^4$~K matching roughly the warm disc phase. This warm phase even becomes predominant around 1~Gyr, and is distinct from the hot phase, making for a bipolar temperature distribution in the outflows.
   At the end of the simulation, we thus find that most of the gas in the galaxy is roughly distributed in four phases: the cold disk around 10~K, a warm phase ($\sim$20 000~K) present in the disc and the outflows, the hot outflow phase ($\sim$7 $\times$ 10$^5$~K), and the very hot halo ($\sim$ 10$^7$~K). We will now investigate how this is related to the presence of cosmic rays and thermal feedback. \\



   \begin{figure}
  \includegraphics[scale=0.53, trim=4 20 0 60, clip]{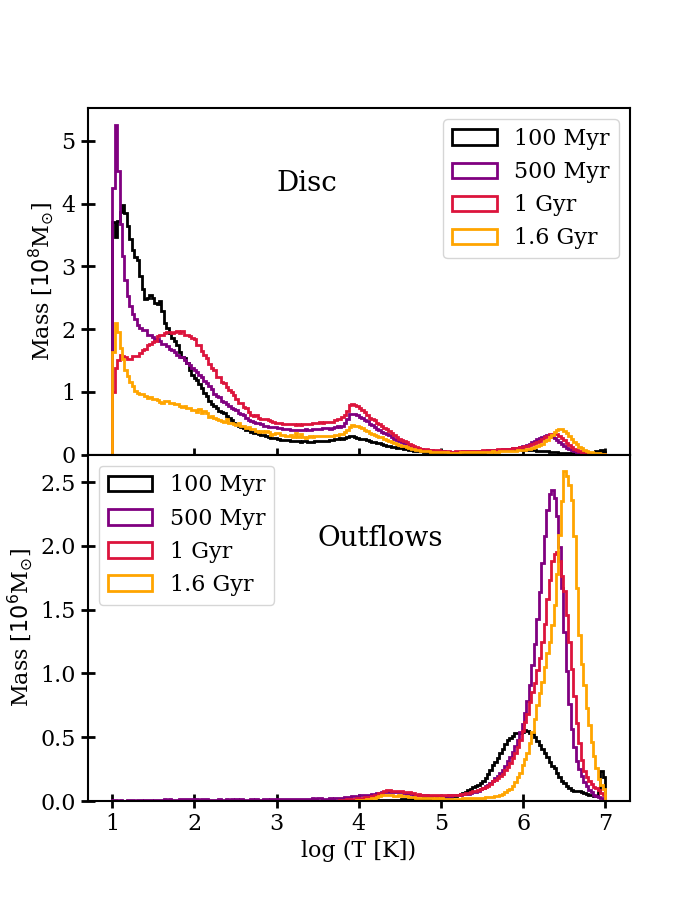}
\caption{Same as Fig. \ref{T_hist_evol}, but for the pt3 simulation (pure thermal). Without cosmic rays, the outflows are predominantly hot.}
  \label{T_hist_evol_pt3}
   \end{figure}

   \subsection{The case without cosmic rays}
   \label{CR_therm}

To understand the role of the thermal feedback and isolate it from the cosmic rays in the evolution of our galaxies and outflows, we will now compare the result of our fiducial simulation cr+t3, to the same simulation run without cosmic ray feedback, pt3 (the pure thermal case). Here again it is important to note that all the results presented below are also valid for the other simulations of our sample. Similarly to cr+t3, we show in Fig. \ref{dens_temp_evol_pt3} the evolution of the pt3 galaxy in terms of gas density and temperatures. 
In terms of gas (and stars) morphology, both simulations have a roughly similar evolution, with the development of spiral arms as well as a bar. However there are some interesting differences: the pt3 simulation has sharper gaseous features, the gas being more concentrated in the spirals and the bar, with little gas in the inter-arm regions and in the rest of the disc. In contrast, the cr+t3 simulation shows more gas outside the spirals, and the features appear more smooth and diffuse. In the edge-on view, one sees that in terms of vertical outflows the evolution is also different. The pt3 simulation clearly pushes less gas into the halo than cr+t3, and most of the outflows stay relatively close to the disc (no further than $\sim$20 kpc altitude). However towards the end of the simulation, it is interesting to note that the outflows seem to start mixing with the halo, making for a smooth gradient of gas density pointing toward the disc accross the entire domain. This is not the case for cr+t3, where the outflows, while propagating through the entire computational domain, still look distinct from the lower-density halo surrounding the outflows.\\
Differences are also visible in the temperature maps: while the spirals and bar concentrate cold gas in both simulations, the cr+t3 also shows a warm gas phase around them which is not present in the pt3 simulation, being replaced by a hot diffuse $\sim$10$^6$~K phase. It is in the edge on views of the temperature maps that the difference is the most striking though. The warm outflowing phase of $\sim$10$^4$~K is completely absent in the pt3 simulation; the outflows appear hotter, with temperatures around 10$^6$~K.
We confirm these findings by measuring the temperature of the disc and the outflows of pt3 in Fig. \ref{T_hist_evol_pt3}. We see indeed in the lower panel that except a very small fraction of gas around 2 $\times$ 10$^4$~K, the outflows are mostly concentrated in one hot phase around 2 $\times$ 10$^6$~K, i.e. higher than even the hot outflowing phase found in cr+t3. In fact the outflowing mass of hot gas seems more important than in cr+t3 (compared to  Fig. \ref{T_hist_evol}). The temperatures in the disc (Fig. \ref{T_hist_evol_pt3}, top panel) are similar to cr+t3, but with the additional presence of gas around 2 $\times$ 10$^6$~K, matching exactly the hot outflow phase. Compared to the four phases found in cr+t3 (cold gas in the disc, warm gas in the disc and outflows, hot outflows and very hot halo, see section \ref{cr+t3}), we therefore find four phases in pt3 as well, but different: the cold disc phase around 10~K, warm gas found only in the disc at $\sim$10$^4$~K, hot gas found in both the disc and the outflows (2 $\times$ 10$^6$~K, hotter than for cr+t3), and of course the very hot halo ($\sim$10$^7$~K). \\
These differences are due to the presence or absence of cosmic ray feedback, since this is the only feature differenciating cr+t3 from pt3. In cr+t3, the cosmic ray injection in supernovae explosions settles a vertical gradient of pressure pointing towards the disc, allowing gas to leave the disc for the halo \citep{1969ApJ...156..445K,1993A&A...269...54B,2012A&A...540A..77D}. This will create outflows made of cooler gas from the disc (the warm phase). On the other hand, the injection of thermal energy into supernova explosions will heat the disc, and blow hot gas into the halo. The corresponding outflows will therefore be composed of hotter gas than the ones only driven by cosmic rays. This explain why we find two phases in the outflows of the cr+t3 simulation, and only one in pt3: while the warm phase is present in the disc in pt3 (Fig. \ref{T_hist_evol_pt3}), there is no mechanism pushing it into the halo in the absence of CRs. Furthermore, the shifting of the hot outflow phase to higher temperatures ($\sim$7 $\times$ 10$^5$~K in cr+t3 vs 2 $\times$ 10$^6$~K in pt3) could be due to adiabatic cooling in the presence of cosmic rays. These results are in agreement with recent work: \cite{2020MNRAS.497.2623J, 2020MNRAS.497.1712B, 2021MNRAS.504.1039R, 2022ApJ...929..170A, 2022arXiv220402410S} also found that the presence of cosmic rays leads to colder gas outflows (mostly also $\sim$ 10$^4$~K), while pure thermal outflows produce higher temperature outflows. Note that while in pt3, a fraction of the hot gas stays in the disc, in cr+t3 almost all of the hot gas phase is evacuated in the outflows from the cumulative effect of cosmic ray and thermal pressure gradients.
Furthermore, the presence of cosmic rays and the corresponding pressure gradient will create diffusion of gas within the disc and oppose gravitational collapse, allowing for a smoother distribution of gas. For the same reasons, the disc is also thinner in the pt3 simulation than in the cr+t3, which the vertical density profiles confirmed, and is in agreement with previous studies \citep{2014MNRAS.437.3312S, 2016ApJ...816L..19G}.\\
In Fig. \ref{outf_sfr_tcr} (top panel) we compare the star formation rate for the cr+t3 and pt3 simulations. We find that cr+t3 shows a lower star formation rate, most likely because the presence of the cosmic ray gradient pressure continuously pushing gas away from the disc leads to a thicker, less dense disc, less prone to very violent star formation phases. This is consistent with recent studies showing star formation suppression by CRs (e.g. \citealt{2020MNRAS.497.2623J,2021ApJ...910..126S}). Furthermore, this pressure gradient added on top of the thermal pressure in cr+t3 also leads to more outflows than in the pure thermal case, as shown in Figure \ref{outf_sfr_tcr} (lower panel), and despite the higher SFR observed for pt3. In this setup, the presence of cosmic rays thus seem more effective than thermal feedback alone at driving large quantities of gas away from the disc into the halo, consistently with previous work \citep{2014MNRAS.437.3312S,2016ApJ...816L..19G,2016MNRAS.456..582S,2018MNRAS.479.3042G,2020A&A...638A.123D,2022MNRAS.513.5000F}. Interestingly, in the first $\sim$100~Myr both the SFR and the OFR of cr+t3 and pt3 show very similar values. This can be linked to the outflow temperatures of Fig. \ref{T_hist_evol}: as mentioned in section \ref{cr+t3}, initially the outflows of cr+t3 are mostly constituted of a hot phase. It therefore appears that initially the outflows in cr+t3 are dominated by the hot phase created by thermal feedback, explaining its similarity with pt3 in its early evolution and outflow production. The cosmic ray pressure gradient then gradually builds up over time, leading to more outflows in a cooler phase (Fig. \ref{T_hist_evol}), and to less star formation, as visible in Fig. \ref{outf_sfr_tcr}.\\

   \subsection{Effect of Angular Momentum input}
   \label{Angmom}

   

\begin{figure}
  \includegraphics[scale=0.57, trim=18 0 0 0, clip]{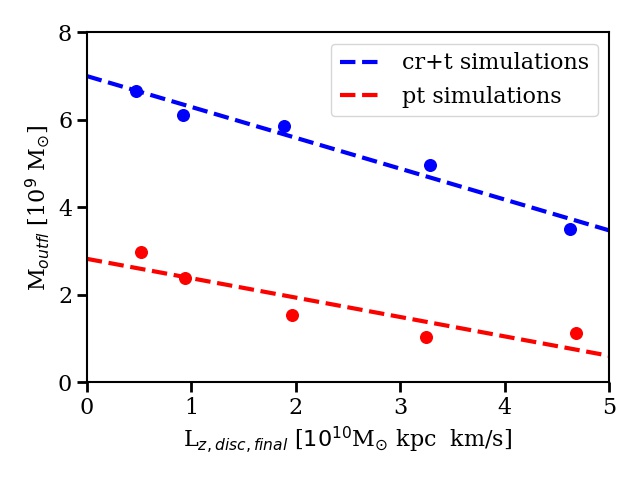}
  \caption{Correlation between the total outflowing gas mass at 10 kpc altitude and the final disc angular momentum for all simulations. We obtain correlation coefficients of $r_{corr}^2$ = 0.96 for the cr+t simulations (CR + thermal), and 0.80 for the pt simulations (pure thermal). With or without cosmic rays, simulations fed with more angular momentum tend to produce less outflows.}
  \label{corr_plot}
\end{figure}

Let us now look at our whole set of simulations, i.e. with different stream inflow velocities, and therefore different amounts of angular momentum (AM) injected into the disc. Just by looking at Fig. \ref{setup}, one sees how much the input of different amounts of angular momentum affects the galaxy evolution. From cr+t1 to cr+t5 (and pt1 to pt5), the disc is more and more extended, as one would expect from higher AM discs \citep{1997ApJ...482..659D}. Differences are very striking in the edge-on view as well: less and less gas is present in the halo from cr+t1 towards cr+t5, especially at higher altitudes, hinting a different outflow rate. In \cite{2021MNRAS.508.4269P}, we had already found that in the presence of CRs, the outflow rate was related to the angular momentum of the disc. More specifically, the total outflowing gas mass was anti-correlated with the angular momentum of the disc, so that simulations with higher stream velocities (higher AM injection) would produce less outflows than lower stream velocities (low AM injection). The reason was that lower AM discs resulting from little AM stream injection were more compact, leading to higher gas densities and a higher SFR. In turn, as described in section \ref{cr+t3}, more star formation produces more CRs, accelerating more gas out of the disc and at higher velocities.\\
To check whether this reasoning still holds in the presence of thermal feedback, we plot in Fig.~\ref{corr_plot} (in blue) the total outflow mass (integrated OFR) at |Z| = 10~kpc as a function of the final gas disc angular momentum, for the 5 simulations cr+t. We see that the anti-correlation is very visible, with a high correlation coefficient ($r_{corr}^2$ = 0.96). Note that the correlation still holds if one looks at different altitudes from the disc. To test whether this correlation is solely the effect of CR feedback, we also check whether it exists in the pt simulations by adding them on the same plot (Fig.~\ref{corr_plot}, in red). We see that even though the scatter is a bit higher ($r_{corr}^2$ = 0.80), we do see a clear decreasing trend of the purely thermal outflows with the disc's angular momentum. It therefore seems that for thermal outflows as well, the outflow rate is directly dependent on the amount of angular momentum in the disc: high AM discs will lead to less outflows, something that is also visible in the bottom two rows of Fig. \ref{setup}. The explanation, similarly to the cosmic ray driven outflows, is that the higher SFR found in low AM discs will lead to more thermal energy injected, accelerating more (hot) gas into the halo due to a higher thermal pressure gradient. The anti-correlation observed between outflow mass and AM for the cr+t simulation is therefore due both to the thermal and CR feedback. Note that due to the differences in inflow velocities, the stream does not reach the disc at the same time for all simulations. We therefore only compute the total outflowing mass from the time the streams hits the disc $t_{hit}$ (determined from visual inspection, ranging from 220 to 310 Myr), and for a total period of 1.39~Gyr for consistency between the different simulations. Consistently, the time at which the final disc angular momentum is computed is $t_{hit}$+1.39~Gyr. \\
It is interesting to note that while both sets of simulations present an anti-correlation of outflows with AM, they do not lie on the same regression line in Fig.~\ref{corr_plot}. This is because, as shown in section \ref{CR_therm}, thermal feedback alone is not able to launch outflows as efficiently as with the presence of CRs; therefore the pt simulations are found at lower gas outflowing masses than the cr+t ones for a given disc angular momentum.

\begin{figure}
  \includegraphics[scale=0.53, trim=4 20 0 60, clip]{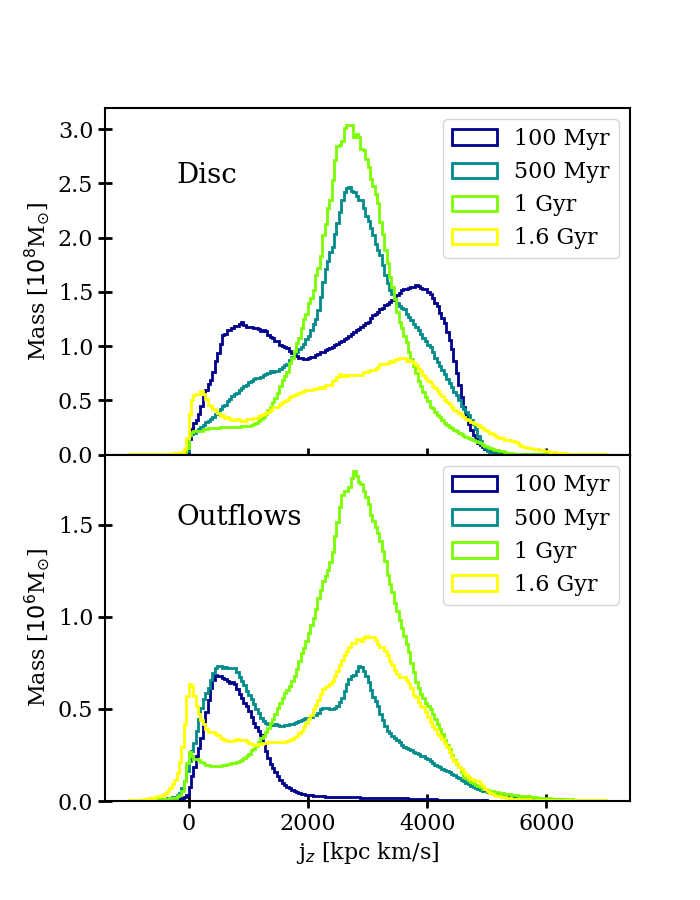}
\caption{Specific angular momentum distribution evolution in the disc (top panel), and in the outflows taken at 5~kpc altitude (lower panel), for the cr+t3 simulation (CR + thermal). For each time the gas temperatures are averaged over 200 Myr, and the outflows are taken within 30~kpc cylindrical radius corresponding to the disc. The peak around 3000~kpc~km/s corresponds to the stream angular momentum injection. Bin size: 40~kpc~km~s$^{-1}$}
  \label{jz_hist_evol}
\end{figure}

\begin{figure}
  \includegraphics[scale=0.56, trim=5 5 0 40, clip]{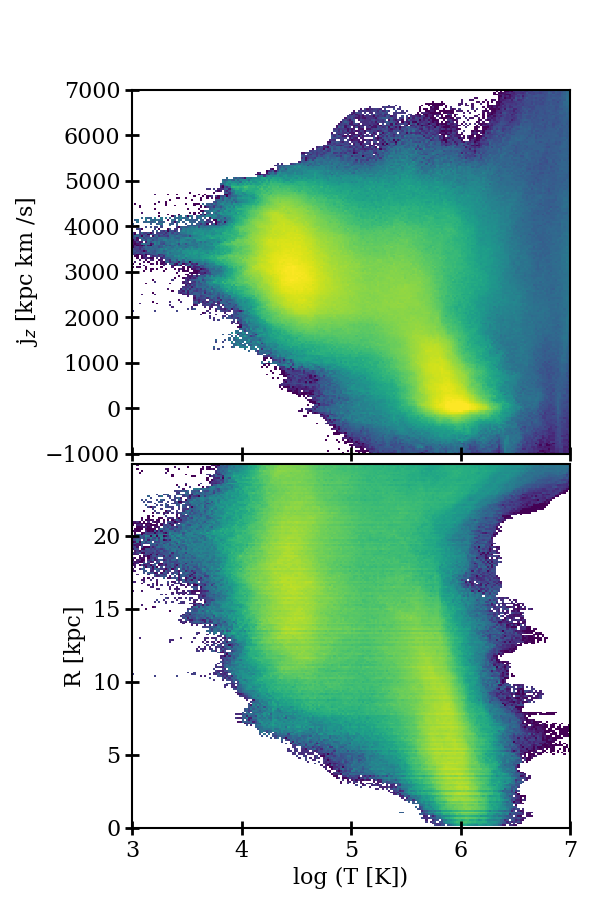}
  \caption{Top panel: specific angular momentum of the outflowing gas as a function of its temperature, taken at 5 kpc altitude in the cr+t3 simulation (CR + thermal), averaged over the last 200~Myr, and within 30~kpc cylindrical radius. Lower panel: cylindrical radius of the outflows as a function of their temperature, for the same gas. We observe two distinct phases in the gas outflows: a hot, low AM phase coming from the centre, and a warm, high AM phase coming rather from the periphery of the disc. Globally there seems to be a decreasing trend of angular momentum with the temperature. }
  \label{T_jz}
\end{figure}

\begin{figure}
  \includegraphics[scale=0.58, trim=18 0 0 0, clip]{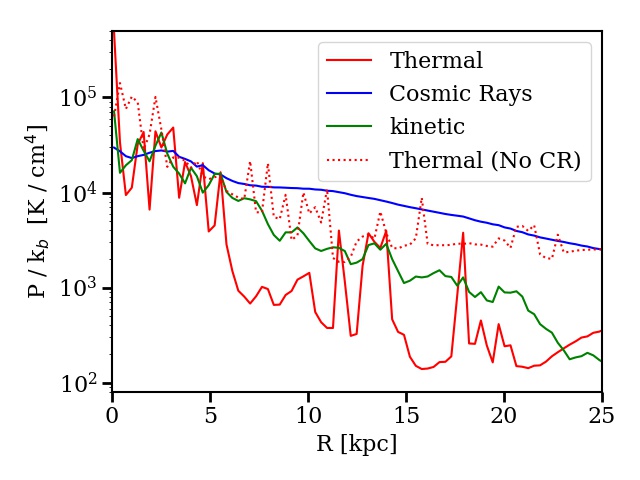}
  \caption{CR, thermal and kinetic pressure as a function of radius in the disc (1~kpc thick) averaged over the last 200 Myr. The solid lines are for the cr+t3 simulation (CR + thermal), and the dotted line for the pt3 simulation (pure thermal). The thermal pressure is significant only in the central region in cr+t3, but is weaker than in the case without cosmic rays.}
  \label{CR_therm_press}
\end{figure}


   \subsection{AM structure and altitude of the outflows}
   \label{structure}

   We show in Fig. \ref{jz_hist_evol} the angular momentum distribution of the disc and the outflows gas, taken at different times. We find that similarly to the outflows temperatures (Fig \ref{T_hist_evol}, lower panel), the AM distribution in the outflows shows a bipolar distribution: initially the outflows are found at low AM, close to 500~kpc~km/s, until high AM gas around 3000~kpc~km/s starts predominating in the outflows. The latter phase is due to the stream accretion; it corresponds to the specific angular momentum of the gas fed into the disc by the stream, which is of 2988~kpc~km/s for cr+t3 (and pt3). This is very visible in the top panel of Fig. \ref{jz_hist_evol} displaying the disc AM distribution, with the growth of a strong peak corresponding to the stream specific AM at 0.5 and 1~Gyr, i.e. as the stream is inflowing into the disc and forms a ring as mentioned in section \ref{cr+t3}. Similarly, in the outflows the corresponding AM phase is strong during this time period, and starts decreasing after the stream accretion time.  We also see that initially the outflows mostly feed from low AM gas, corresponding to the central region of the galaxy, before the stream allows strong outflows at higher AM. We do observe this bipolar AM distribution for the other cr+t simulations as well, but with the high AM phase shifted to the left or the right depending on the AM stream input. The input of high AM gas from the stream is consistent with previous studies showing the entry of high AM streams into galaxies \citep{2015MNRAS.449.2087D}.\\
   In Figure \ref{T_jz} (top panel) we display the outflows specific angular momentum of the outflows as a function of their temperature. We see that even though the gas displays large ranges of temperatures and AM, the hot phase described in section \ref{cr+t3} (around 10$^6$~K) shows a high concentration for very low AM, while the warm phase is composed mostly of higher AM gas, from roughly 2000 to 4000~kpc~km/s. Furthermore, even within those two phases, there seems to be a decreasing trend of AM with temperature: the hotter the gas, the lower its AM. To understand this, we display in the lower panel of Figure \ref{T_jz} the cylindrical radius of the outflows as the function of their temperature. The plot shows clearly that a high amount of hot gas is coming from the very center, explaining its low AM. In fact at low radii the gas is always hot, while the gas from the periphery shows predominantly cooler temperatures (constituting the warm phase). This can be understood by looking at the thermal and cosmic ray pressure as a function of cylindrical radius in the disc, shown in Fig. \ref{CR_therm_press}: the thermal feedback is most efficient in the inner $\sim$6 kpc, which is where the star formation density is the highest (especially towards the end of the simulation), and indeed where we find the hot outflows originating from. Globally the thermal pressure decreases quickly as a function of the cylindrical radius, quicker than the CR pressure. This is very consistant with e.g. \cite{2017MNRAS.465.4500P} and \cite{2022ApJ...929..170A} (in a different setup with postprocessing of a stratified disc), who show that the ratio of cosmic ray to thermal pressure decreases for increasing star formation density and decreasing radius. Therefore as one moves to the periphery of the galaxy, thermal feedback becomes less efficient and outflows are mostly launched by the cosmic ray gradient of pressure, pushing cooler gas into the halo. This is why the hot outflows have predominantly low AM, and explains the decreasing trend of AM with temperature observed in Fig. \ref{T_jz} (top panel). \\
         \begin{figure}
  \includegraphics[scale=0.55, trim=15 0 0 0, clip]{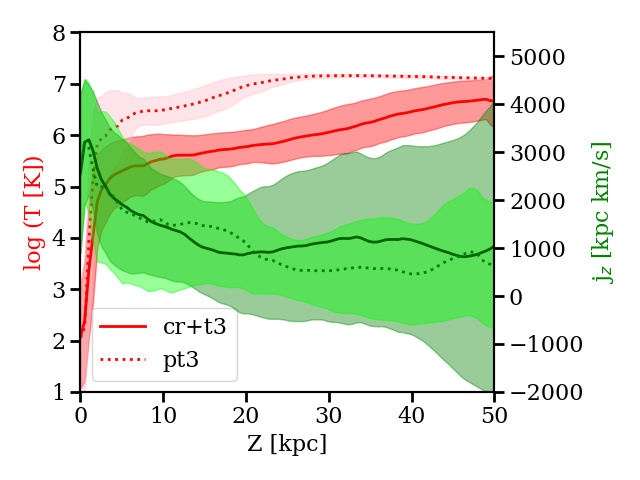}
\caption{Mean temperature (in red) and angular momentum (in green) of the outflows as a function of vertical distance from the disc, with their 1-$\sigma$ dispersion, at t = 1.7~Gyr, and inside a 30~kpc radius cylinder. The solid lines are for the cr+t3 simulation (with CRs), and the dotted lines for pt3 (without). The 1-$\sigma$ dispersions are indicated in lighter colors for the pt3 simulation. As one moves towards higher altitudes, outflows have predominantly lower angular momentum and higher temperatures.}
\label{meanTjz}
\end{figure}
         \begin{figure*}
  \includegraphics[scale=0.5, trim=70 30 0 50, clip]{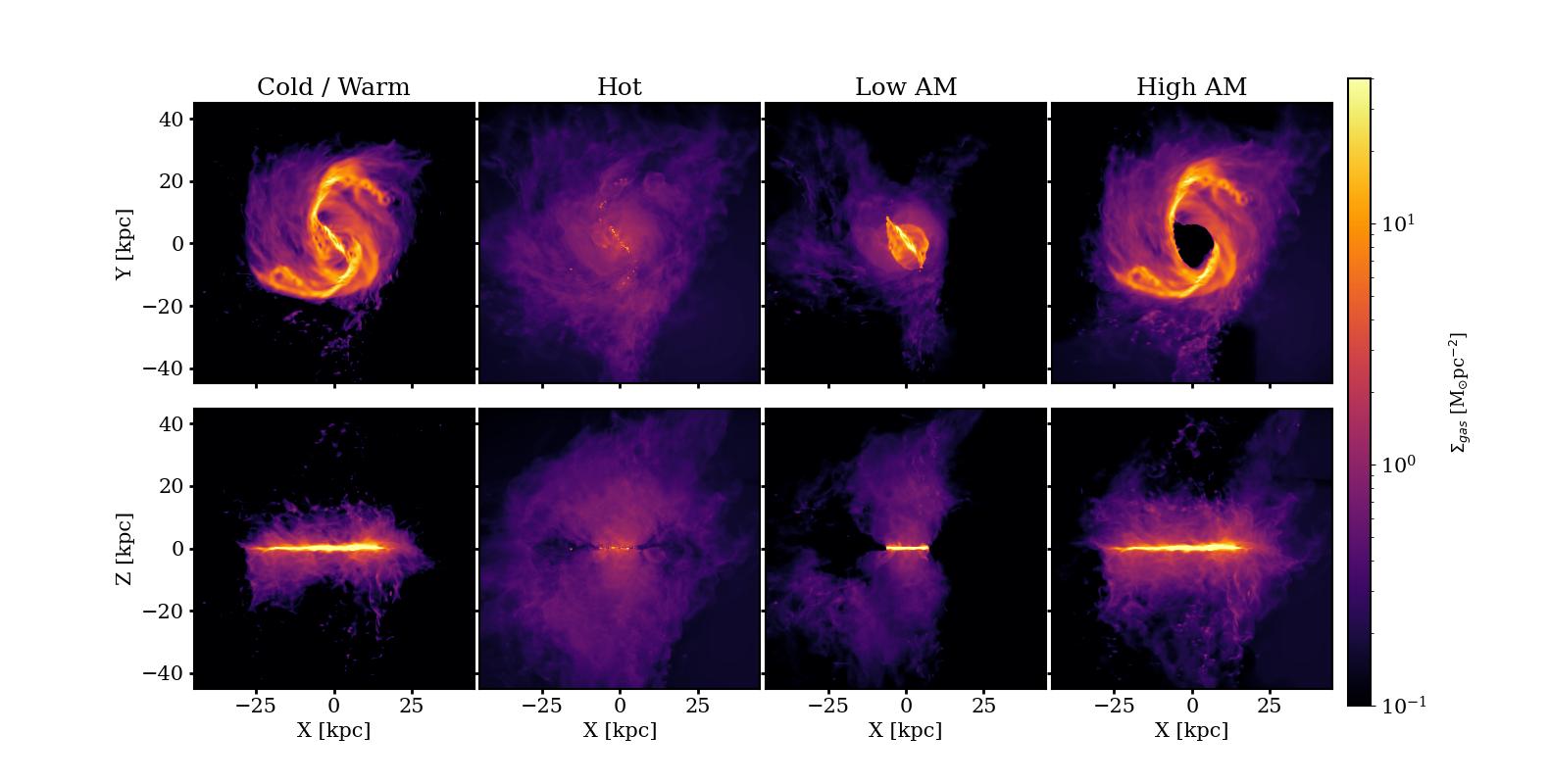}
\caption{Gas column density of cr+t3 (CR + thermal) at t=1.7 Gyr, for 4 different phases: the gas below 10$^5$~K (cold and warm gas), the gas above 10$^5$~K (hot), the low angular momentum gas (j$_z<$1500~kpc~km/s), and the high AM gas (j$_z>$1500~kpc~km/s). There is an overlap of the cold/warm and high AM phases, as well as of the hot and low AM ones. High altitude outflows are mostly hot and with low AM, while warm and high AM gas stay in the vicinity of the disc. }
\label{gasphases_cr+t3}
         \end{figure*}
%
    To visualize these different gas phases, we map in Fig~\ref{gasphases_cr+t3} the column density of gas divided into four categories: low and high temperatures, as well as low and high AM, based on the bipolar distributions we found for these observables. This figure highlights a number of interesting results. First, although not identical we do see strong overlaps of the cooler gas phase with the high AM one, as well as the hot phase with the low AM one, consistently with Fig. \ref{T_jz} (top panel). As already shown, the hot phase seems to be mostly originating from the disc central region, therefore coinciding with the low AM phase. Furthermore, we see that the hot outflows are the only ones reaching altitudes up to $\sim$40~kpc and above, while the cooler outflows stay in the vicinity of the disc (below  $\sim$20~kpc altitude). In terms of AM, although we do find a bit of high AM gas at higher altitude, the low AM phase also appears more efficient at penetrating deep into the halo. We confirm these results by plotting in Fig. \ref{meanTjz} the mean temperature and mean specific angular momentum of the outflowing gas as a function of altitude, showing a clear decrease for the AM and a clear increase of the temperature. We therefore find again this link between AM and temperature here, and see that the outflows at high altitudes are mostly composed of hot gas with low AM emanating from the center, while the high AM gas with low temperatures tends to stay close to the disc. \\
    %
    To understand these trends of the outflows temperature and AM with altitude in the light of thermal and CR feedback, let us turn to the pt3 simulation. As can be seen in Fig. \ref{jz_hist_evol_pt3}, we also find two phases of AM in the outflows, as expected from the stream input feeding high AM to the disc. In fact both the disc and the outflows specific angular momentum distributions are very similar to the ones found for cr+t3, except for the outflows high AM peak which tends to be at lower masses. The situation is different here though since as shown in section~\ref{CR_therm}, we do not have the warm phase in the outflows. By plotting the AM and cylindrical radius as a function of the temperature in Fig.~\ref{T_jz_pt3} we find that the hot phase is constituted of gas at all angular momenta and all radii, instead of having this high concentration at high AM and low temperatures as for cr+t3. The thermal pressure as a function of radius in the disc of pt3 is plotted alongside the thermal and CR pressure of cr+t3 in Fig. \ref{CR_therm_press}. It shows that the presence of CRs reduces the thermal pressure in the disc, consistently with previous work, e.g. \cite{2022MNRAS.517..597C}. This is likely due to the lower SFR found in simulations with cosmic rays, leading to lower thermal energy injection. The higher thermal pressure explains why we found the hot outflows to be more important in pt3 than cr+t3 in section \ref{CR_therm}, and, together with the adiabatic cooling in the cosmic ray case, why the hot phase is shifted to higher temperatures in pt3.
    Displaying again the different gas phases maps in Fig~\ref{gasphases_pt3}, we see a strong bipolarity in temperature: the cold and warm gas are found only in the disc, while the gas present in the halo is always hot. This confirms that only the presence of CRs is able to push cooler gas into the halo. We show in Fig. \ref{meanTjz} the average temperature and angular momentum of the outflows as a function of altitude alongside the ones of cr+t3, and find a similar trend: the angular momentum of the outflows decreases as one moves away from the disc, while the temperature increases (and is higher at all altitudes than in the presence of cosmic rays). \\
    %
    We already showed in section \ref{CR_therm} that the presence of CRs created stronger outflows due to the combination of thermal with CR pressure pushing warm gas away from the disc. Additionally, we find here that the presence of CRs launches a hot, predominantly low AM outflowing gas from the central region of the galaxy, which will be removed from the disc and is able to penetrate into the halo at altitudes not reachable from thermal driven outflows alone. This will have a direct impact on the star formation in the centre, and could explain the suppression of central thermal pressure, as well as the generation of weaker hot outflows from there. Furthermore, as originating from the galaxy centre, these outflows are likely metal-rich, so that cosmic rays should prove to be an efficient way at transporting metal enriched gas deep into the metal-poor halo. The result of CRs allowing higher altitude outflows has also been found in recent studies (e.g. \citealt{2021MNRAS.501.3640H}), and can be explained again by the superposition of both the thermal and the cosmic ray pressure gradient: hot gas is accelerated more from its thermal pressure than cold gas, and after leaving the disc is still accelerated by the CR pressure gradient, which is absent in the case of purely thermal feedback. We show in Fig. \ref{T_vz_hist} that indeed higher velocities are reached for higher gas temperatures in the outflows, therefore pushing hot gas to higher altitudes. This is very consistent with \cite{2022MNRAS.517..597C}, who find that in presence of CRs the hot gas is the fastest and leaves the disc vicinity, while the cooler gas is in fountain flows. However, to our knowledge this is the first time angular momentum composition is added to this picture. Note that we find this relation between vertical velocities and temperatures for the pt simulations as well, explaining why we obtain similar trends of temperatures and angular momentum with altitude. However in that case the velocities all decrease very quickly as one moves to higher altitudes, because of the absence of a cosmic ray pressure gradient to support the outflows after they leave the disc vicinity. \\
    Therefore, in the presence of both thermal and CR feedback, the global picture is as follows. Higher star formation density in the center (especially at later times in our setup) leads to hot outflows dominating in the central region of the galaxy because this is where thermal pressure is the highest. These central hot outflows have low AM and thanks to their high thermal pressure gradient combined with the CRs pressure gradient, can be accelerated to very high velocities and altitudes. In contrast, at the periphery the thermal pressure is low and cosmic rays are the dominant mechanism pushing gas out of the disc; therefore the corresponding high AM outflows tend to be cooler, and thus stay in the vicinity of the disc. Note that these results are also mostly valid for the other cr+t and pt simulations; we find very similar temperature phases, with just the high angular momentum phase shifted depending on the stream angular momentum. We do find some quantitative differences though: in particular for the low AM input runs, the increased star formation rates pushes hot gas velocities to even higher velocities and altitudes, something that can be seen in Fig. \ref{setup}. However qualitatively everything described in this section remains true, and we do observe that even for the lowest angular momentum runs cr+t1 and pt1 producing the most outflows, the latter remain confined within 25-30~kpc for pt1, while they fill the whole computational domain for cr+t1 (first column of Fig. \ref{setup}). \\




         \begin{figure}
  \includegraphics[scale=0.53, trim=4 20 0 60, clip]{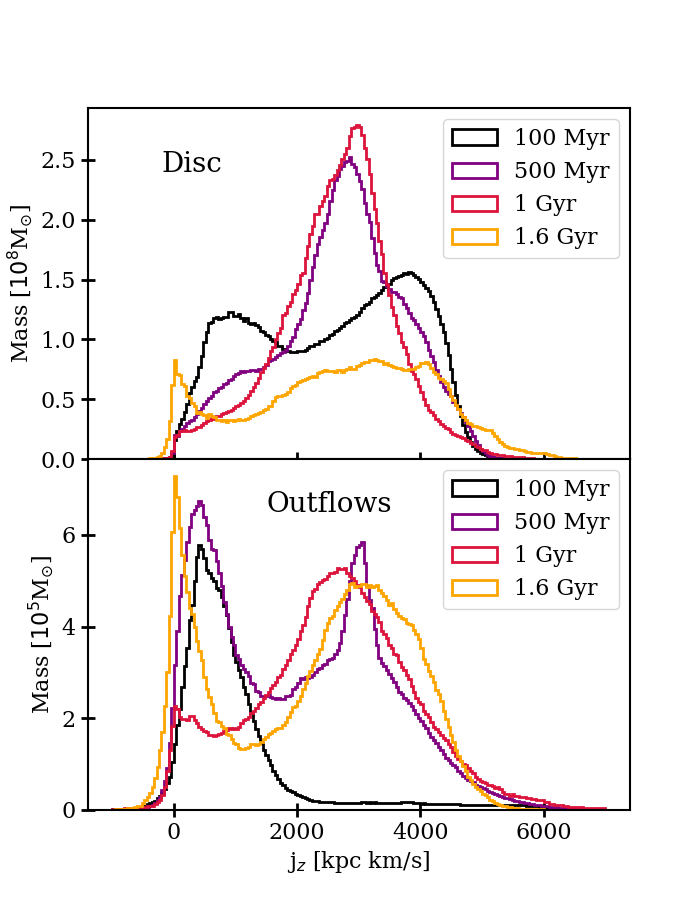}
\caption{Same as Fig. \ref{jz_hist_evol}, but for the pt3 simulation (pure thermal).}
  \label{jz_hist_evol_pt3}
         \end{figure}
\begin{figure}
  \includegraphics[scale=0.56, trim=5 5 0 40, clip]{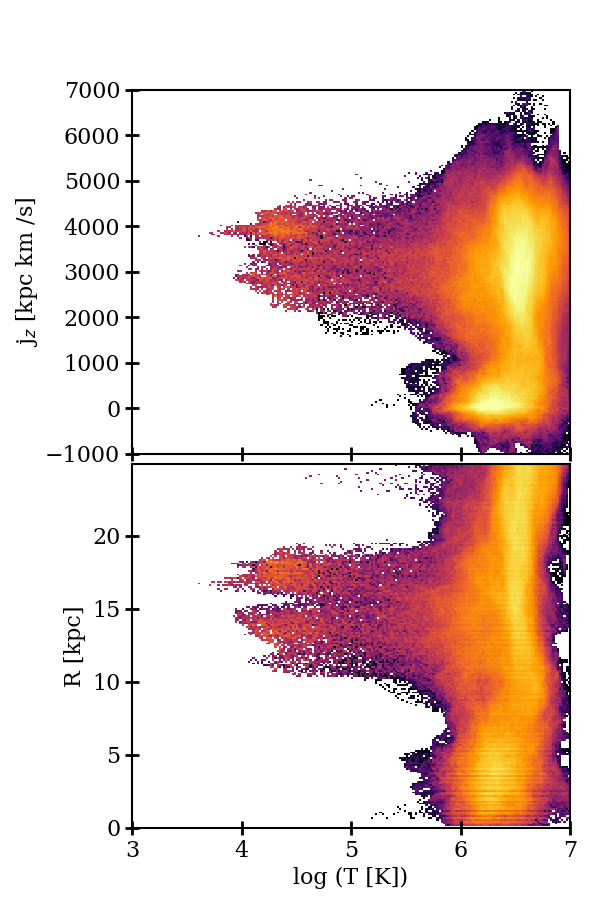}
  \caption{Same as Fig.~\ref{T_jz}, but for the pt3 simulation (pure thermal).}
  \label{T_jz_pt3}
\end{figure}
\begin{figure*}
  \includegraphics[scale=0.5, trim=70 30 0 50, clip]{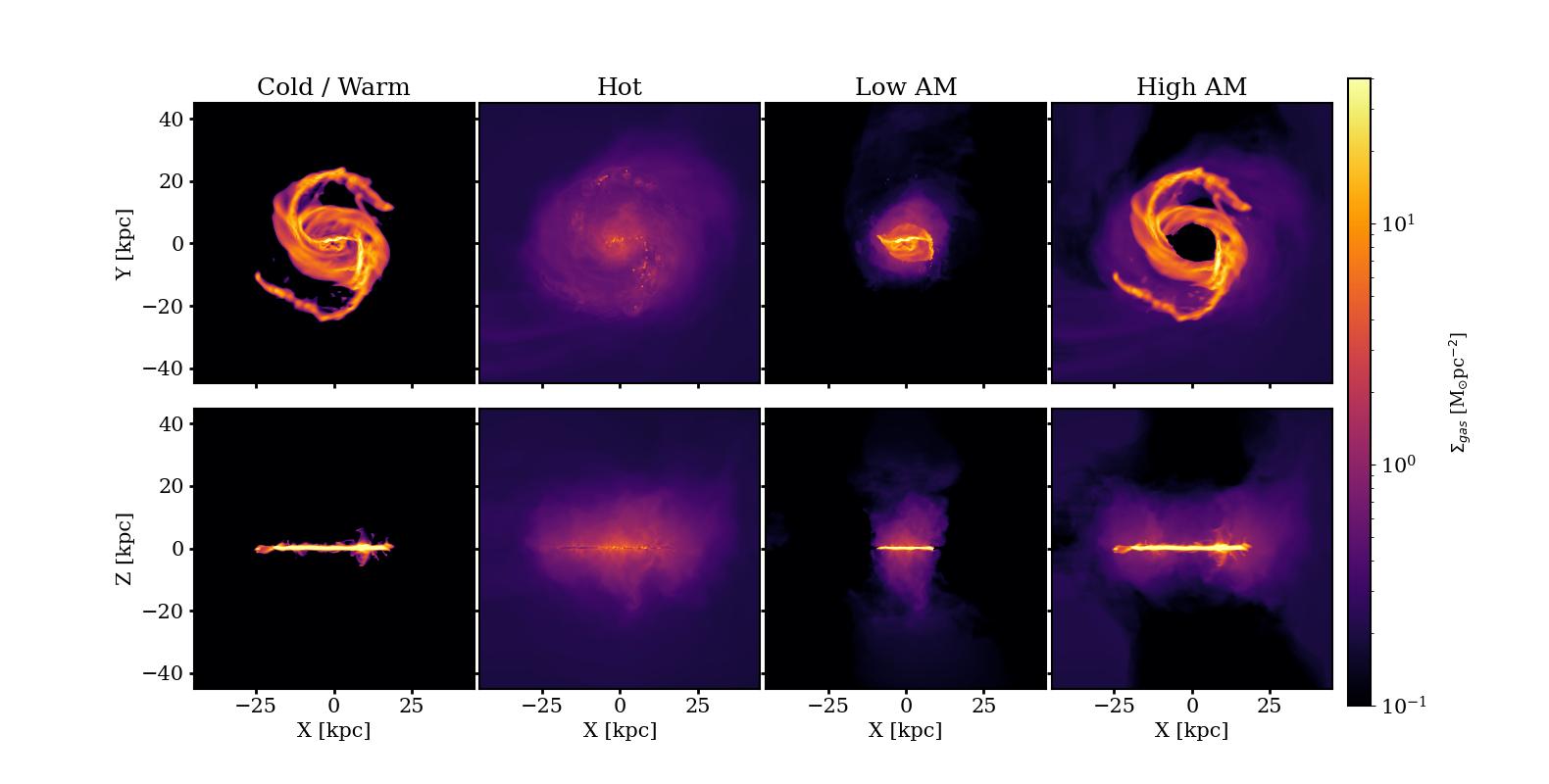}
\caption{Same as Fig.~\ref{gasphases_cr+t3}, but for the outflows of the pt3 simulation (pure thermal). In the absence of cosmic rays, most of the outflows stay close to the disc, and cold and warm gas do not leave the disc.}
\label{gasphases_pt3}
\end{figure*}

    \section{Discussion}
    \label{discussion}

   Stream feeding has been shown to be an important aspect of galaxies formation, and our work shows the outflows generation and properties in this context. The stream accretion leads to a few interesting results; as we showed in section \ref{structure}, it leads to a strong high angular momentum gas phase in the disc, which is then found in the outflows as well. This leads to these two very distinct phases of hot, low AM outflows, and warm, high AM ones, with the former being the only one reaching high altitudes. To check whether this picture still holds without the stream accretion, we also ran simulations without stream, by just letting the disc evolve on its own. We find that in this setup, as expected, less star formation happens at the periphery of the galaxy, since the ring formed by the stream is not present. Therefore initially most of the star formation happens in the centre, leading to dominating hot, low AM outflows. Within this phase we do observe a decreasing trend of AM with temperature, as shown with the stream as well. However over time a second, higher AM phase develops in the outflows as well, showing that star formation is not present only in the centre anymore. This then leads to results very similar to the ones shown in the presence of the stream, with the difference that due to the absence of the star forming ring, the warm, high angular momentum phase is smaller. We also compared this case to a simulation without cosmic rays (and without stream), and found that consistently with the rest of this study, the outflows stay closer to the disc, and are weaker for pure thermal feedback. Therefore, even without the stream continuously feeding the galaxy with higher AM gas, the results presented in this paper seem to remain qualitatively valid. \\   
     Some recent studies \citep{2020MNRAS.497.2623J,2021MNRAS.501.3640H} found that the presence of cosmic rays tends to suppress the outflows in the vicinity of the disc rather than increasing them. While we found similar results in terms of high altitude outflows and outflows temperatures in the presence of cosmic rays, in contrast to these studies we have higher outflow rates with cosmic rays than without. We do see a small reduction of the hot outflow phase from cosmic rays, but which is largely compensated by warm gas outflows, absent in the pure thermal case. In fact by looking at the outflow rate in the midplane (altitude 0), we do find slightly more outflows in the case without cosmic rays (similarly to Fig. 4 of \cite{2021MNRAS.501.3640H}), but as soon as one looks above (or under) the disc the trend is inverted, because the cosmic rays take over the thermal pressure very quickly outside the disc. We specify that this is not linked to the stream feeding, as outflows already start dominating in the case with cosmic rays before the stream has reached the disc, and we observe this trend even without the stream inflow (see paragraph above). In contrast to these papers, our work therefore tends to show enhancement of the outflows by CRs out of the disc even at low altitudes, which is in agreement with many other studies about cosmic rays such as \cite{2014MNRAS.437.3312S,2016ApJ...816L..19G,2016MNRAS.456..582S,2018MNRAS.479.3042G,2020A&A...638A.123D,2022MNRAS.513.5000F}.\\
   While implementing the new stellar feedback module in \textsc{piernik}, we tried different configurations that led us to adopt our final setup. In particular, we ran simulations without the implementation of the momentum kick prior to stellar feedback energy injection, and compared them to the case with kick. While the differences were relatively modest in the presence of cosmic rays, in the pure thermal feedback simulations we observed a substancial reduction of the outflow rate without the presence of momentum kick; by the end of the simulation there was almost no more gas pushed higher than a few kpc above and below the disc. This is most likely due to too rapid cooling of the injected gas in the dense environement of the disc as mentioned in section \ref{thermal}, and confirms the efficiency of the initial kick in resolving this issue. We faced a similar problem when trying to implement a feedback scheme based on \cite{2017ApJ...843..113B}, which includes gradual energy injection in star forming regions. Injecting energy in small amounts at each timestep led to suppression of the outflows due to too rapid cooling. \\
   In \cite{2021MNRAS.508.4269P}, we showed the anti-correlation between outflow rates and disc angular momentum for cosmic-ray driven winds. The setup used in the present paper is similar (with the addition of thermal feedback), but with a few key distinctions that lead to quantitative differences in the galaxy evolution and outflows. The entry of the stream into the disc is smoother here because of the stream being wider and less concentrated; therefore the important SFR peak observed in \cite{2021MNRAS.508.4269P} when the stream hit the disc is not present here, or at least much weaker. In terms of angular momentum, the implementation of the stellar particle creation tends to deplete the gaseous disc, reducing his mass and therefore total angular momentum. This explains why the final disc angular momentum values tend to be lower than in \cite{2021MNRAS.508.4269P}, where we did not have gas removal from star formation. The depleting of the disc mass, especially after the end of the stream feeding, is visible in Fig. \ref{T_hist_evol} and \ref{T_hist_evol_pt3} (top panels). One further difference, significant for the outflows production, is the use of isotropic cosmic ray diffusion, in contrast to the anisotropic diffusion used previously, as described in section \ref{piernik}. This explains why despite general qualitative agreements, the outflow rates and mass loadings found in \cite{2021MNRAS.508.4269P} were higher; in the isotropic case cosmic rays leave the disc more easily, making the CR pressure gradient weaker. We believe nonetheless that the isotropic setup is more realistic than the anisotropic one at this resolution, for the reasons detailed in  section \ref{piernik}. In fact many studies on Cosmic Rays have successfully used isotropic CR diffusion \citep{2013ApJ...777L..16B,2014MNRAS.437.3312S,2020MNRAS.497.2623J}, so that our results should be at least qualitatively correct. However we note that the use of anisotropic diffusion at higher resolution would be an improvement of this model; we then expect the outflows to be stronger, possibly accentuating the gap between the pure thermal and cosmic ray + thermal feedbacks cases. \\
   The stream feeding used in this paper is an idealised setup where the gas stream arrives in the disc plane. We showed in \cite{2021MNRAS.508.4269P} simulations where the stream inflows from above the disc. We found that all the results in terms of angular momentum were still holding, with the notable difference that due to the more violent impact of the stream into the disc (particularly as the stream used in these simulations was narrower), the star formation rate and therefore outflows productions was lower. Since all the results were qualitatively similar however, for the work shown in the present paper we kept only the idealised in-plane stream inflow. \\
   


\begin{figure}
  \includegraphics[scale=0.57, trim=15 0 0 20, clip]{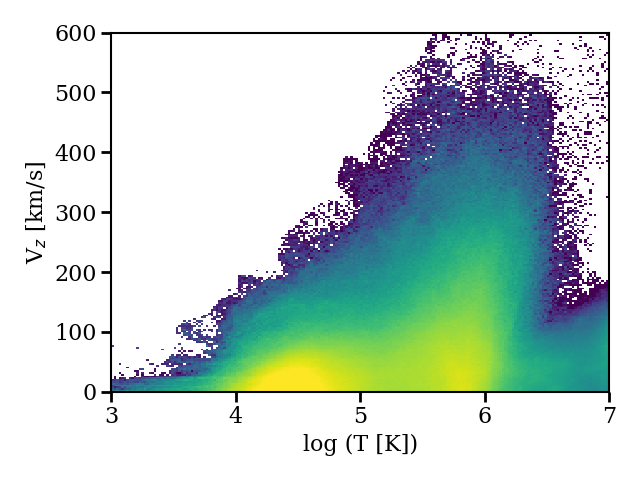}
  \caption{Vertical velocities of the outflows taken at 5 kpc altitude as a function of their temperature, averaged over the last 200 Myr of the cr+t3 simulation (CR + thermal) and within 30~kpc cylindrical radius. Higher velocities are reached for the hottest outflows only.}
  \label{T_vz_hist}
\end{figure}




\section{Summary and Conclusions} 
\label{ccl}

In this paper, we showed the impact of cosmic ray stellar feedback together with thermal feedback on galaxy evolution and on the wind production, by running MHD simulations of disc galaxies including cosmic rays, as well Nbody stellar and dark matter particles. These simulations were performed thanks to new additions to the \textsc{piernik} code \citep{2010EAS....42..275H, 2010EAS....42..281H, 2012EAS....56..363H, 2012EAS....56..367H} presented here for the first time, namely cooling and heating of the thermal gas, a new star formation module including thermal and cosmic ray energy injections as well as stellar particle creation, and the use of a Riemann solver. We also simulated stream feeding to the galaxy, an important galaxy growth mechanism predicted by cosmological simulations \citep{2009Natur.457..451D} and linked to the presence of galactic rings \citep{2015MNRAS.449.2087D}. This was done by implementing a constant external inflow of cool gas from the intergalactic medium for 1~Gyr (for a total simulated time of 1.7~Gyr), bringing mass and angular momentum to the disc. By running different simulations, each with a different stream inflow velocity, we were able to change the amount of angular momentum injected into the disc, and compare the link between angular momentum and outflow production. Furthermore, we ran a set of simulations including only thermal feedback, and another with both cosmic rays and thermal feedbacks, to highlight the effect of cosmic rays. The latter build up a gradient of pressure allowing gas to leave the disc, and creating strong outflows \citep{1993A&A...269...54B, 2008ApJ...674..258E,2014ApJ...797L..18S}. We list our results below: \\

\noindent $\bullet$ The inclusion of both cosmic ray and thermal feedback leads to strong outflow production in all cases, which are able to leave the galactic disc and reach high altitudes of several dozens of kpc. \\
$\bullet$ We find an anti-correlation between the disc angular momentum and the outflowing mass both with and without cosmic rays. In other words, the more angular momentum the stream brings to the disc, the weaker the outflows. This is due to low angular momentum discs having a more clumpy, higher density distribution, and therefore more star formation enhancing the pressure gradients of both the thermal and the cosmic ray components. \\
$\bullet$ The pure thermal feedback cases lead to weaker outflows than the simulations including cosmic rays. \\
$\bullet$ The presence of cosmic rays suppresses star formation, as well as thermal pressure in the disc, due to a thickening of the disc and gas loss to the halo. \\
$\bullet$ The outflowing gas in the presence of cosmic rays shows two temperature phases, a hot ($\sim$10$^6$~K) and a warm phase ($\sim$20 000~K), while without cosmic rays the outflows are only hot ($\sim$2 $\times$ 10$^6$~K). This is due to cosmic rays being able to push cooler gas away from the disc, while thermal feedback alone only acts on hot gas. Furthermore, the hot outflow phase in presence of cosmic rays is slightly reduced, and cooler. \\
$\bullet$ The gas in the outflows also shows two distinct phases in angular momentum: a low angular momentum phase from gas ejected in the central region of the disc, and a high angular momentum phase coming from the ring formed by the stream accretion. This is found in both pure thermal simulations and in simulations with cosmic rays. \\
$\bullet$ We find a link between the specific angular momentum of the outflowing gas and its temperature in the presence of cosmic rays. Specifically, we observe in the outflows a hot, low angular momentum gas phase, distinct from a warm, high angular momentum one. Overall there seems to be a decreasing trend of the specific angular momentum of the gas with its temperature. We explain this by the higher star formation density found in the central region of the galaxy, where the thermal pressure is thus higher and leads to hotter outflows. As one moves away from the center, thermal energy becomes less efficient than cosmic rays at driving gas away from the disc, and the cosmic ray pressure gradient allows cooler gas to leave the disc.\\
$\bullet$ We also find that these two distinct gas phases are found at different distances from the disc. While the warm gas with high angular momentum tends to stay in the vicinity of the disc, the hot gas with low angular momentum is the only one able to escape the disc and reach high altitudes ($\sim$40~kpc and above). This is likely due to the combination of thermal and cosmic ray pressure gradients, which accelerates hot gas to higher velocities than the warm one. Cosmic rays thus seem efficient at removing predominantly hot and low angular momentum gas from the central region of the disc, where the metallicity is expected to be higher. This will have an impact on the star formation in the centre, and can transport large amount of metal enriched gas deep into the metal-poor halo, impacting directly the composition of the circumgalactic medium. \\
$\bullet$ In the pure thermal feedback case, without the cosmic ray pressure gradient able to accelerate the gas after leaving the disc, the gas stays closer to the disc and does not seem to be able to escape it. \\

\noindent Therefore, we show strong evidences that the presence of cosmic rays dramatically changes the global picture of winds production and evolution, in terms of outflow rates, temperatures, angular momentum distributions and altitude. This, together with previous work on this topic (e.g. \citealt{2012MNRAS.423.2374U,2013ApJ...777L..16B,2016ApJ...816L..19G,2020MNRAS.492.3465H,2022MNRAS.513.5000F}), indicates cosmic rays to be an important component to be included in galaxy simulations. \\
In future work, we plan to examine the presence of fountain flows in this setup, and their dependence and impact on the different gas phases described in this paper. 

\vspace{1cm}
\noindent \textbf{AKNOWLEDGEMENTS} \\

\noindent This work was supported by the Polish National Science Center under grant 2018/28/C/ST9/00443.
Calculations were carried out at the Centre of Informatics - Tricity Academic Supercomputer \& networK (TASK) and on the HYDRA cluster at the Institute of Astronomy of Nicolaus Copernicus University in Toruń. Michał Hanasz acknowledges  support by the National Science Center through the OPUS grant No. 2015/19/B/ST9/02959. Thorsten Naab acknowledges support by the Excellence Cluster ORIGINS which is funded by the DFG (German research foundation) under Germany’s Excellence Strategy – EXC-2094 – 390783311.
We thank Varadarajan Parthasarathy for his contribution in the implementation of HLLD Riemann solver.

\vspace{1cm}
\noindent \textbf{DATA AVAILABILITY} \\

\noindent  The data underlying this article will be shared on reasonable request to the corresponding author.

\bibliographystyle{mnras}
\bibliography{biblio}


\appendix 
\section{Tests of the Riemann solver} \label{sect:appendix}

\subsection{Brio-Wu sock tube}

\begin{figure}
  \includegraphics[scale=0.5, trim = 30 80 0 20]{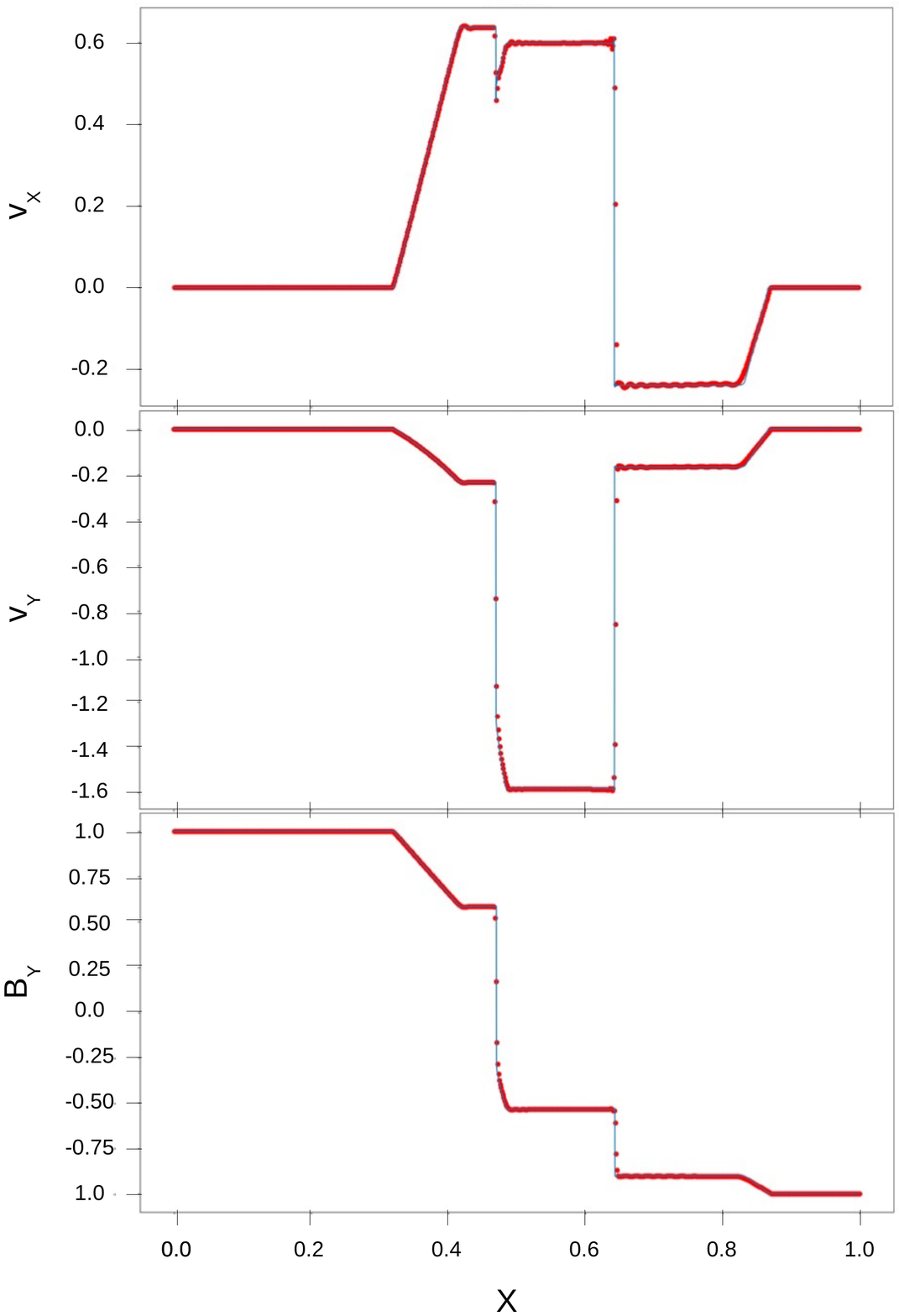}
  \caption{Results of the Brio and Wu test, obtained with the HLLD solver of {\sc piernik} code, for $t=0.1$ and resolution of 800 grid cells (points) and 20000 cells (full line). Velocity components $v_X$ and $v_Y$ are shown in two upper panels, magnetic field components $B_Y$ is shown in the lower panel. Results are consistent with the original results. See Figure 2 in \citet{1988JCoPh..75..400B} and Figure 5a in \citet{1995ApJ...442..228R}.}
  \label{fig:brio-wu}
\end{figure}

One of the standard 1D test problems of MHD is the Brio and Wu test problem \citep{1988JCoPh..75..400B}.
The initial conditions for left and right states assume: 
$\rho_L=1$, $p_L=1$, $v_{xL}=0$, $v_{yL}=0$, $B_{xL}=0.75$, $B_{yL}=1$,
$\rho_R=0.125$, $p_R=0.1$, $v_{xR}=0$, $v_{yR}=0$, $B_{xR}=0.75$, $B_{yR}=-1$ and adiabatic index $\gamma=2$.
Results of the Brio and Wu test shown in Figure~\ref{fig:brio-wu} demonstrate the correct operation of the HLLD Riemann Solver implemented in the \textsc{piernik} code.

\subsection{2D MHD blast wave}

\begin{figure}
  \centerline{ \includegraphics[scale=0.41, trim=10 70 0 70, clip]{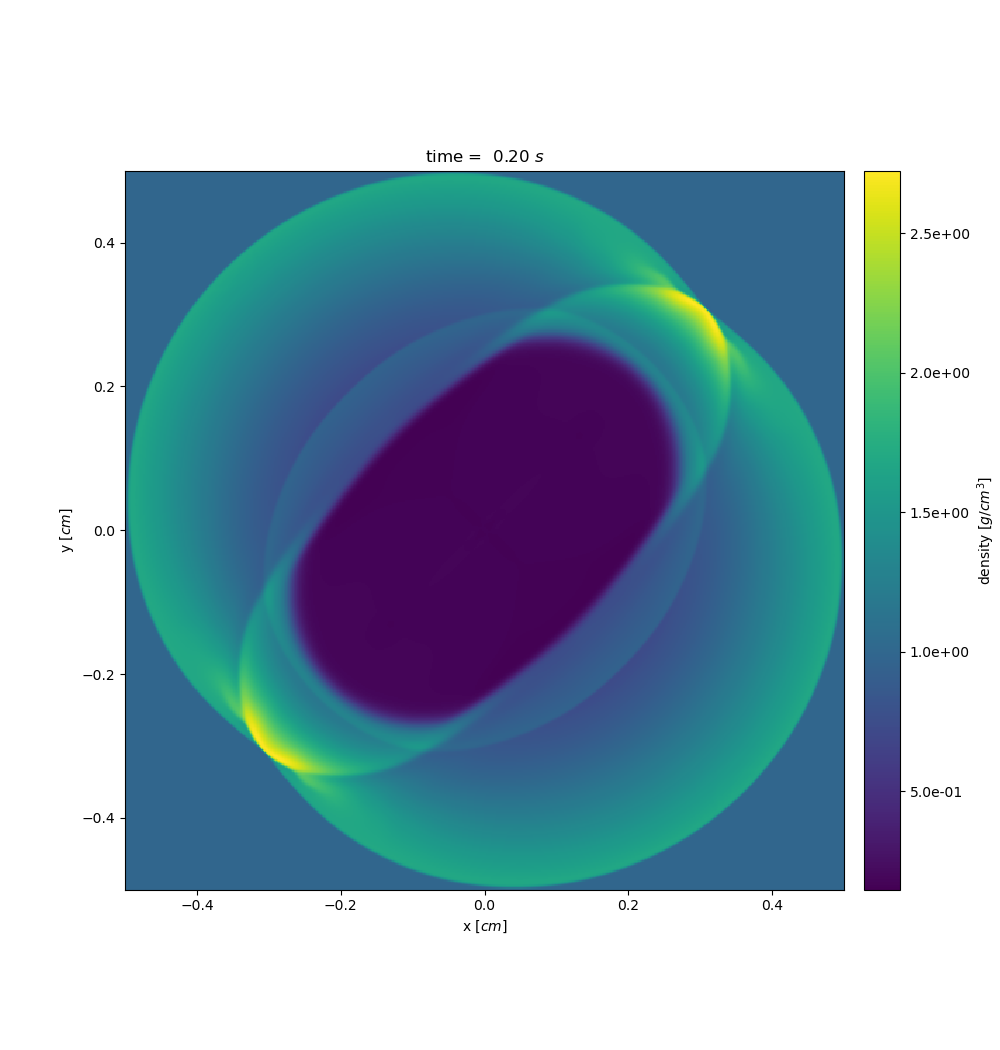}}
  \caption{Gas density distribution for the 2D MHD blast wave test.
 }
  \label{fig:mhd-blast}
\end{figure}

In Figure~\ref{fig:mhd-blast} we show the results of the magnetized Sedov  blast wave problem.
This test is useful in investigating the capacity of the solver at handling the grid alignment effects, and preserving the
symmetry of the out-going blast wave. We performed a 2D MHD blast wave test with the HLLD solver, using the magnetic divergence cleaning algorithm by \cite{2002JCoPh.175..645D}. The initial density in the computational volume $\rho_{\rm ext}$ equals $1.0$ and the pressure $p_{\rm ext}$ equals 0.1. Initial velocities are zero everywhere. At $t = 0$, the uniform magnetic field is initialized with $B_x = B_y =  1/2$. 
The blast is initiated within the region $r \leq 0.1$, where the pressure is 10, i.e. 100 times the ambient pressure. 
We use a square domain $ -0.5 \leq x, y \leq 0.5 $  in $512 \times 512$ grid points. The geometry of the resulting shock-wave pattern agrees with corresponding structures resulting from other  MHD codes \citep[see eg.][]{2005JCoPh.205..509G,2016MNRAS.455...51H}.

\end{document}